\documentclass[desactivate]{aa}
\usepackage[utf8]{inputenc}
\usepackage{float}
\usepackage{placeins}
\usepackage{graphicx}
\usepackage{hyperref}
\usepackage{color}
\usepackage{subfigure}
\usepackage[T1]{fontenc} 
\usepackage{linenoaa}

\usepackage{txfonts}

\newcommand{\Msun}{\,\mathrm{M}_\odot}

\newcommand{\tablecontinuum}{
\begin{table} 
\caption{Continuum Fit Parameters and Results}             
\label{tab:disk}      
\centering
\begin{tabular}{ p{4.5cm}|p{3cm}}
 \hline
Parameters  & Values \\
 \hline
 Inclination i$_D$ & 86.6 $\pm$ 0.1$^\circ$ \\
 Inner radius  $R_{Din}$  & $< 4$ au \\ 
 Outer radius  $R_{Dout}$& 317 $\pm$ 5 au\\
 T$_{Do} =$ T$_D({100\,\mathrm{au}})$ & 9.8 $\pm$ 0.6 K \\
 Exponent of $T_D$ & -0.3 $\pm$ 0.2 \\
 $\Sigma_{Do}$ = $\Sigma_D({100\,\mathrm{au}})$ & $1.6\pm0.3\,10^{23}$\,cm$^{-2}$ \\ 
 Exponent of $\Sigma_D$ & 0.48 $\pm$ 0.04 \\
Dust emissivity index $\beta$  & 0.63 $\pm$ 0.07  \\
Absorption coefficient $\kappa_0$ & $0.052$\,cm$^{2}/g$ \\ \\
 \hline
 H$_{Do}$ = H$_D({100\,\mathrm{au}})$ & 11.6 au\\ 
 Exponent of H$_D$(r), $h_d$ & -1.38  \\
  \hline
\end{tabular}
\tablefoot{\small{
Dust emissivity is $\kappa(\nu) = \kappa_0 (\nu/345\,\mathrm{GHz})^\beta$. H$_{Do}$ and $h_D$ were fixed to the best fit values found from a simultaneous fit of the 220 and 330 GHz data only.}
}
  \label{tab:obtained-cont}
\end{table} 
}

\newcommand{\tablemodel}{
\begin{table*}[!th] 
\small
\caption{Best fit models for molecular lines}             
\centering    
\begin{tabular}{|c c|l|l|l|}        
\hline              
\multicolumn{2}{|c|}{Parameters} &  \multicolumn{1}{c|}{$^{12}$CO 2-1 \& 3-2} & \multicolumn{1}{c|}{$^{13}$CO 2-1 \& C$^{18}$O 2-1} &\multicolumn{1}{c|}{HCO$^+$ 3-2} \\ 
\hline  
$\delta$V & (km\,s$^{-1}$) &  $0.20 \pm 0.01$  & $0.15 \pm 0.01$ & 0.15 \\
$T_{atm}(r)$ & (K) & $31.2 \pm 0.2\,(r/R_0)^{-0.68 \pm 0.02}$ & $16.8 \pm 0.2\,(r/R_{0})^{0.07 \pm 0.01}$ & $15.0 \pm 0.2\,(r/R_{0})^{(-0.12 \pm 0.03)}$ \\
$T_{mid}(r)$ & (K)&  $\min(T_{atm},7.4 \pm 0.2\, (r/R_{0})^{-0.27})$ & $\min(T_{atm},10.4 \pm 0.1\, (r/R_{0})^{(-0.27 \pm 0.03)}$ & $\min(T_{atm},9.1 \pm 0.4\,(r/R_{0})^{-0.27}$ \\ 
$H(r)$= & (au) &$ 27.9 \pm 0.2\,(r/R_{0})^{1.16 \pm 0.02}$ & $24.0 \pm 0.2\,(r/R_{0})^{1.25 \pm 0.03}$ & $30.8 \pm 0.8\,(r/R_{0})^{0.93 \pm 0.13}$\\
$\Sigma(r)$ & (cm$^{-2}$) & $8.2 \pm 1.2 \, 10^{18}\,(r/R_{0})^{-3.1 \pm 0.2}$ & $^{13}$CO : $4.5 \pm 0.4\,10^{17}\,(r/R_{0})^{-3.3 \pm 0.2}$ &$ 6.1\pm 1.0 \,10^{13}\,(r/R_{0})^{-1.6 \pm 0.9}$ \\
 & &  & C$^{18}$O : $1.9 \pm 0.1\,10^{16}\,(r/R_{0})^{-3.3 \pm 0.2}$ &\\  
$\Sigma_{dep}$   & (cm$^{-2}$) & $3.7\pm0.4\,10^{22}$  &  $3.5\pm0.1\,\,10^{22}$ & $8.7\pm0.2\,\,10^{22}$ \\
\hline 
Inclination & ($^\circ$) & $85.5 \pm 0.6$ & $89.8 \pm 0.4$ & $87.4 \pm 0.8$\\
R$_{in}$ & (au) &  61 $\pm$ 1 & 77 $\pm$ 2 & 53 $\pm$ 1\\
R$_{out}$ & (au) & 760 $\pm$ 5 & 790 $\pm$ 6 & 521 $\pm$ 2\\
 \hline 
\end{tabular}

\label{tab:gas-model}
\tablefoot{Error bars are 1$\sigma$ formal errors. All parameters are defined in Appendix \ref{app:modeling}. Parameters with no error bars were fixed. The distance used is 160\,pc and the V$_\mathrm{LSR}$ is 7.35 km\,s$^{-1}$.  
The mid-plane temperature exponent being loosely constrained, we fixed its value to that found from $^{13}$CO 2-1 for all lines. Following Eq.\ref{Temperature}, we fix the exponent $\delta$ that controls the steepness of the gradient to 2.5, see Appendix \ref{app:modeling}.  
The pivot $R_0$ is fixed at 100 au to maximize the sensitivity. For all lines, we checked that the disk is in Keplerian rotation.} 
\end{table*} 
}

\newcommand{\figtomo}{
\begin{figure*}
  \subfigure(a){\hspace{5cm}} 
  \subfigure(b){\hspace{5cm}} 
  \subfigure(c){}
  \centering

  \subfigure{\includegraphics[width=0.268\textwidth]{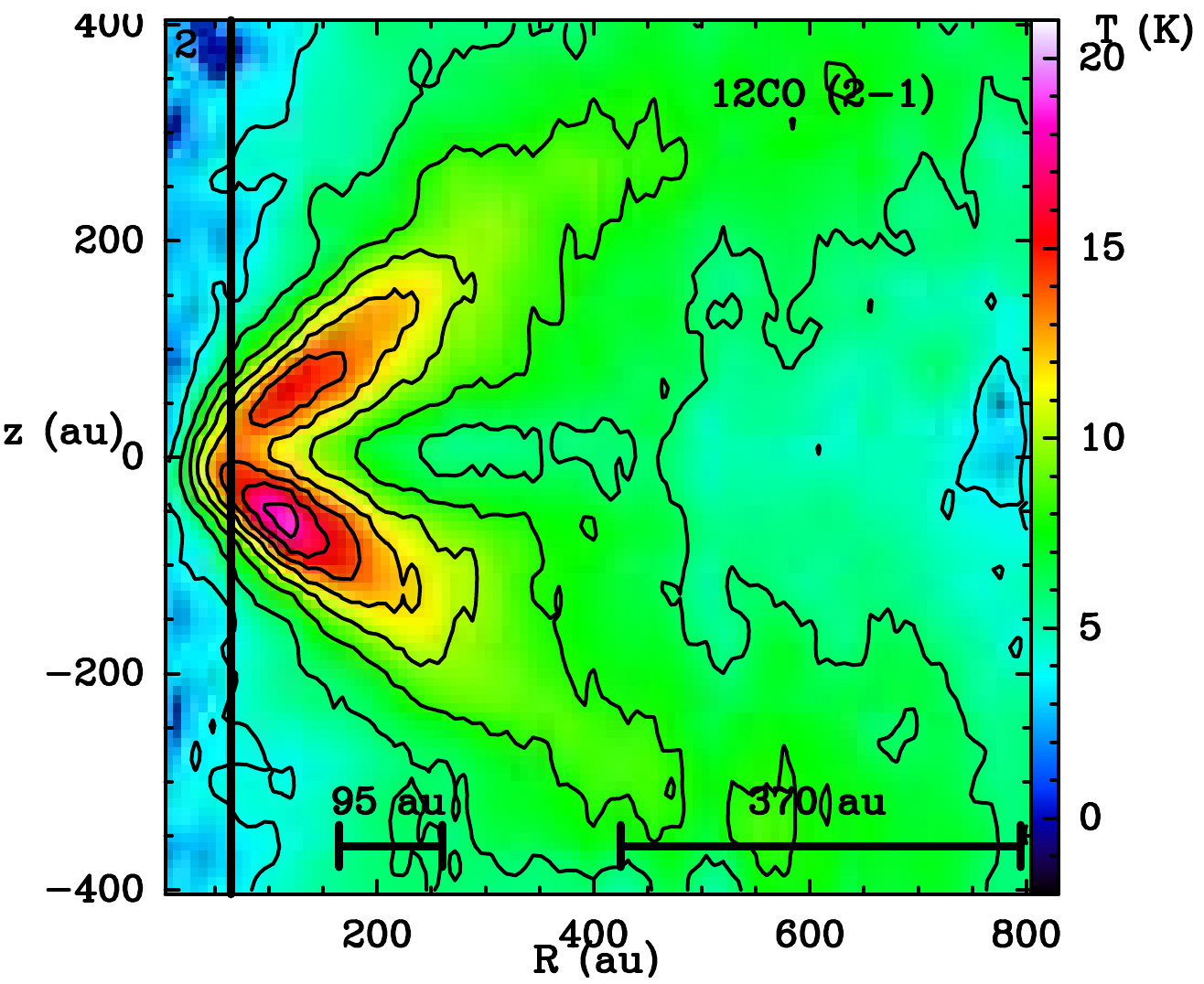}} 
  \subfigure{\includegraphics[width=0.268\textwidth]{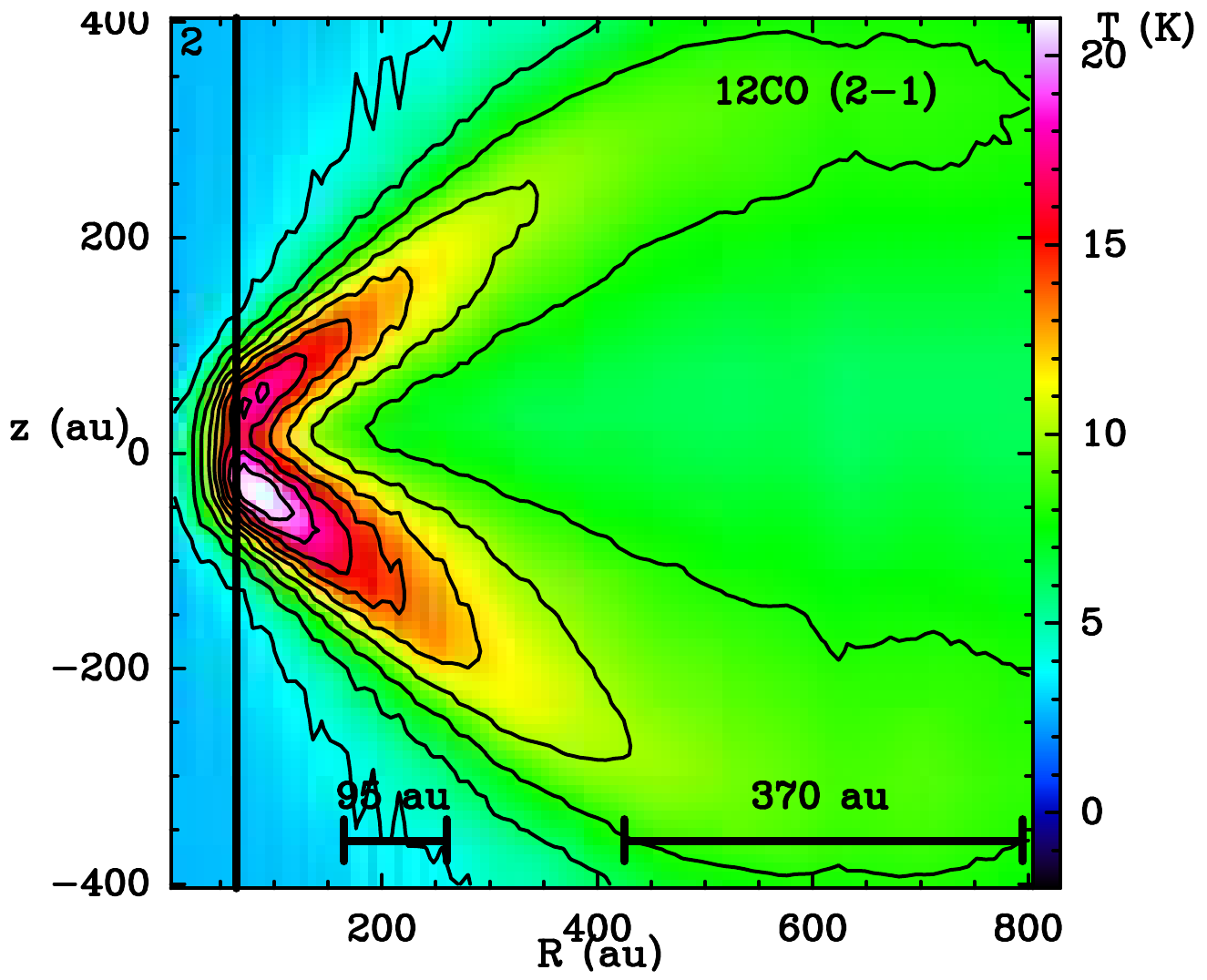}} 
  \subfigure{\includegraphics[width=0.268\textwidth]{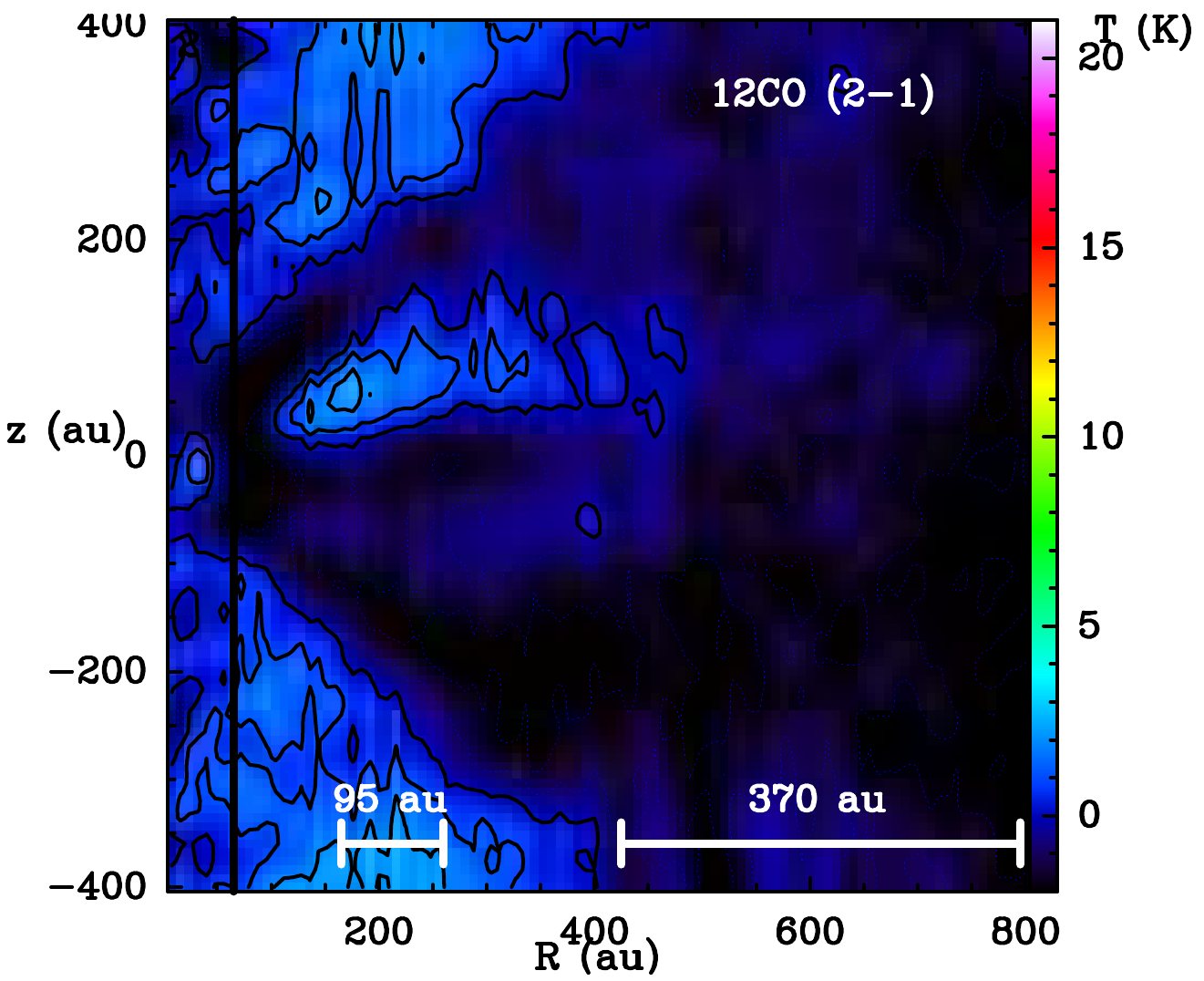}}
  
  \centering
  \subfigure{\includegraphics[width=0.268\textwidth]{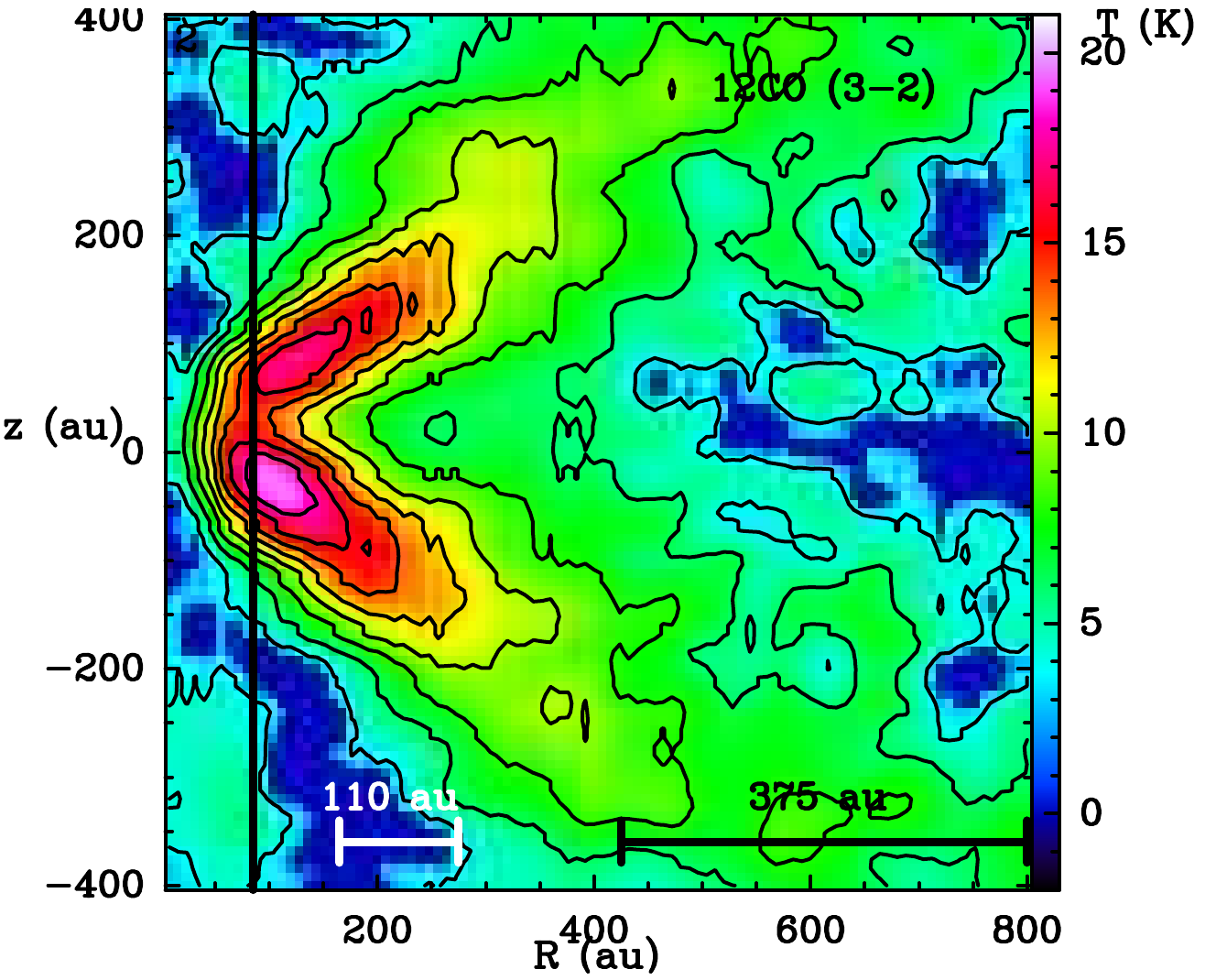}} 
  \subfigure{\includegraphics[width=0.268\textwidth]{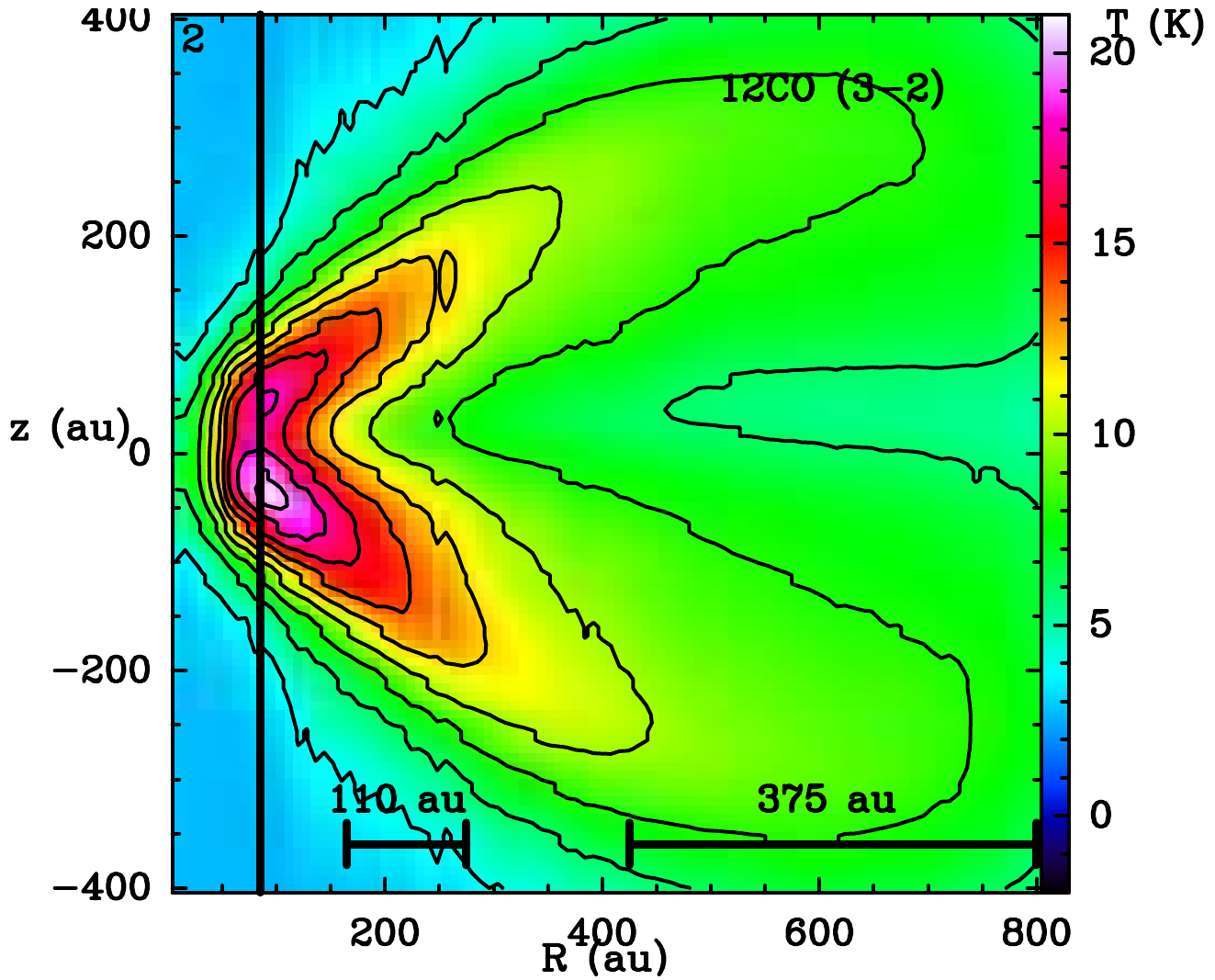}} 
  \subfigure{\includegraphics[width=0.268\textwidth]{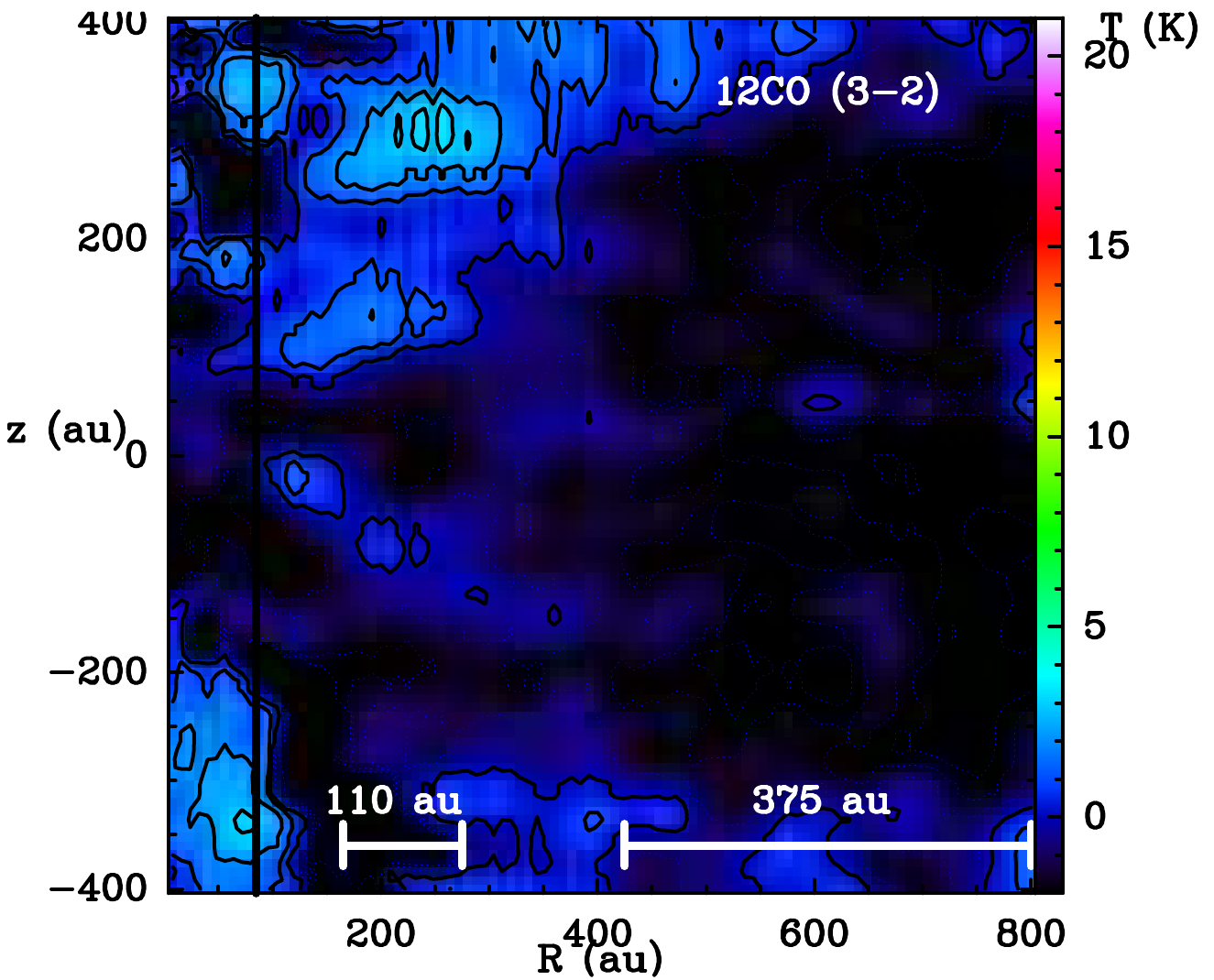}}

  \centering
  \subfigure{\includegraphics[width=0.268\textwidth]{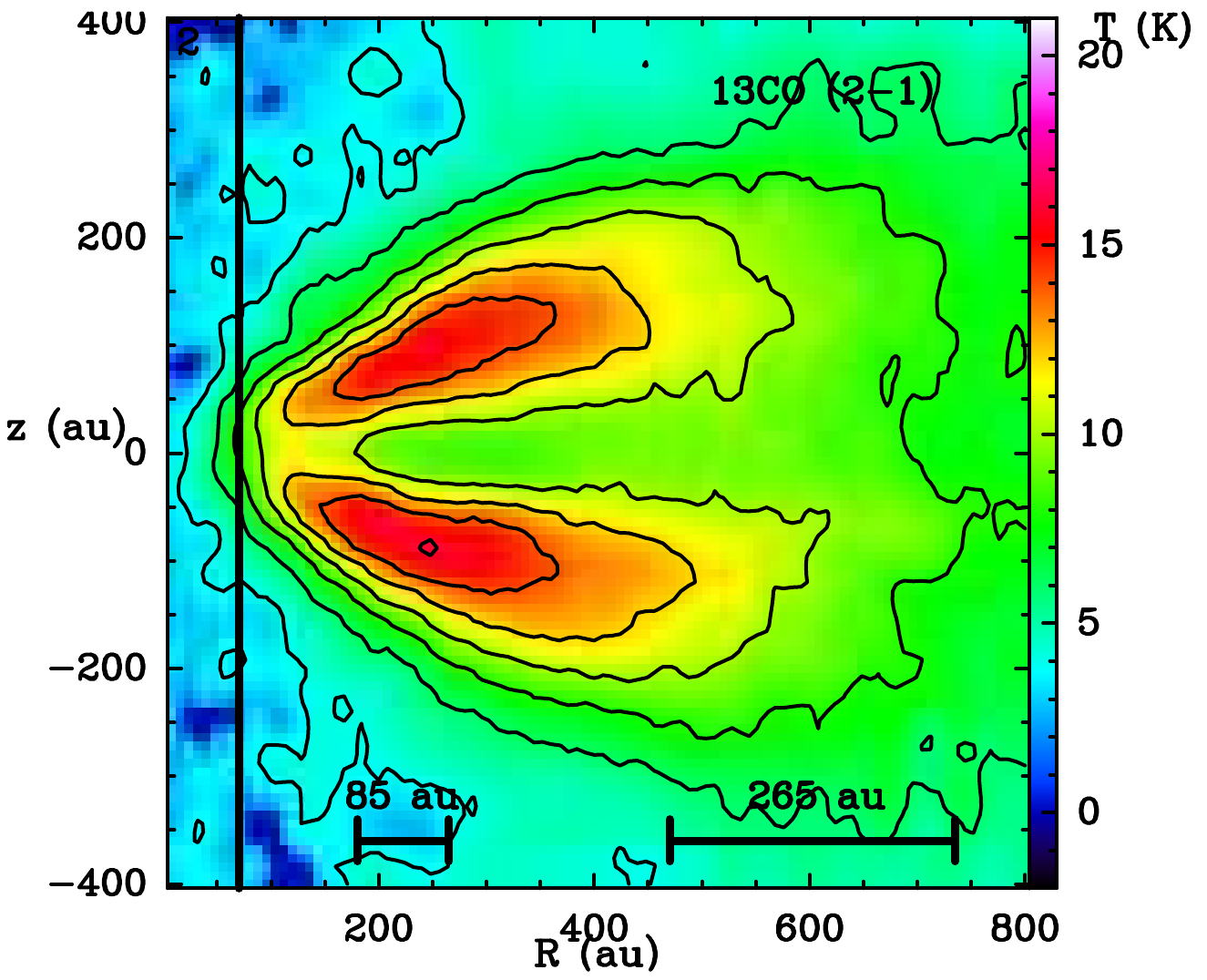}} 
  \subfigure{\includegraphics[width=0.268\textwidth]{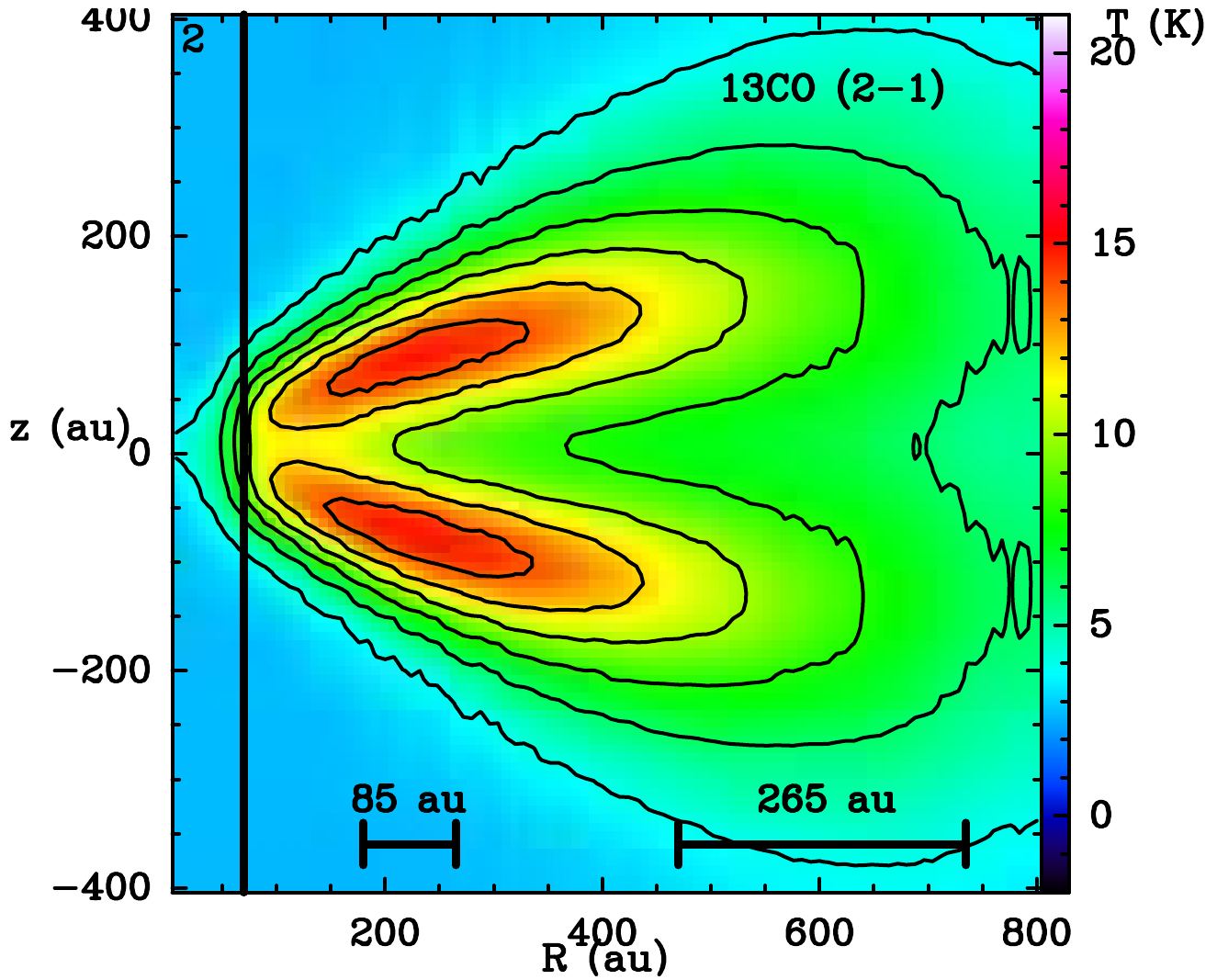}} 
  \subfigure{\includegraphics[width=0.268\textwidth]{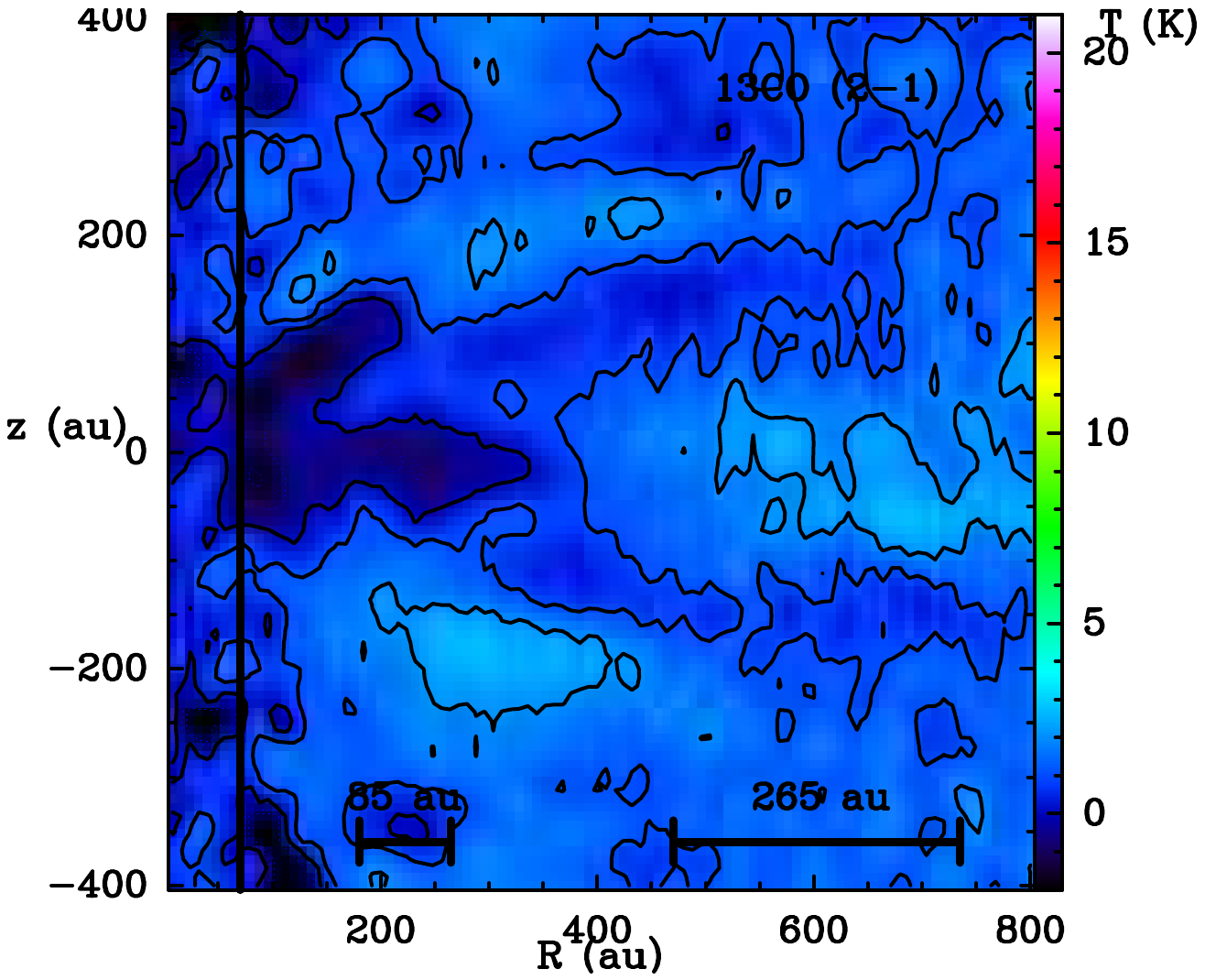}}
  
  \centering
  \subfigure{\includegraphics[width=0.268\textwidth]{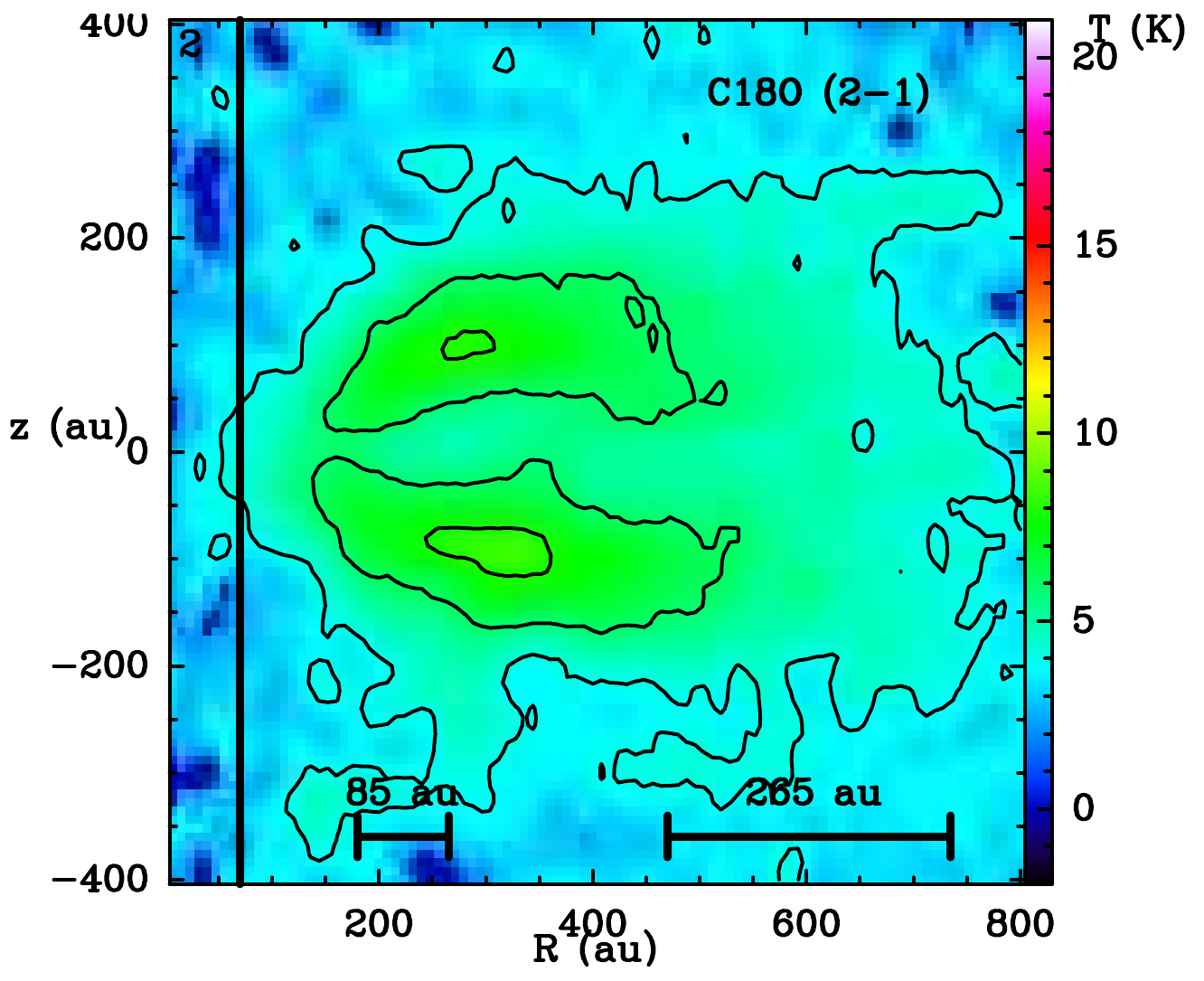}} 
  \subfigure{\includegraphics[width=0.268\textwidth]{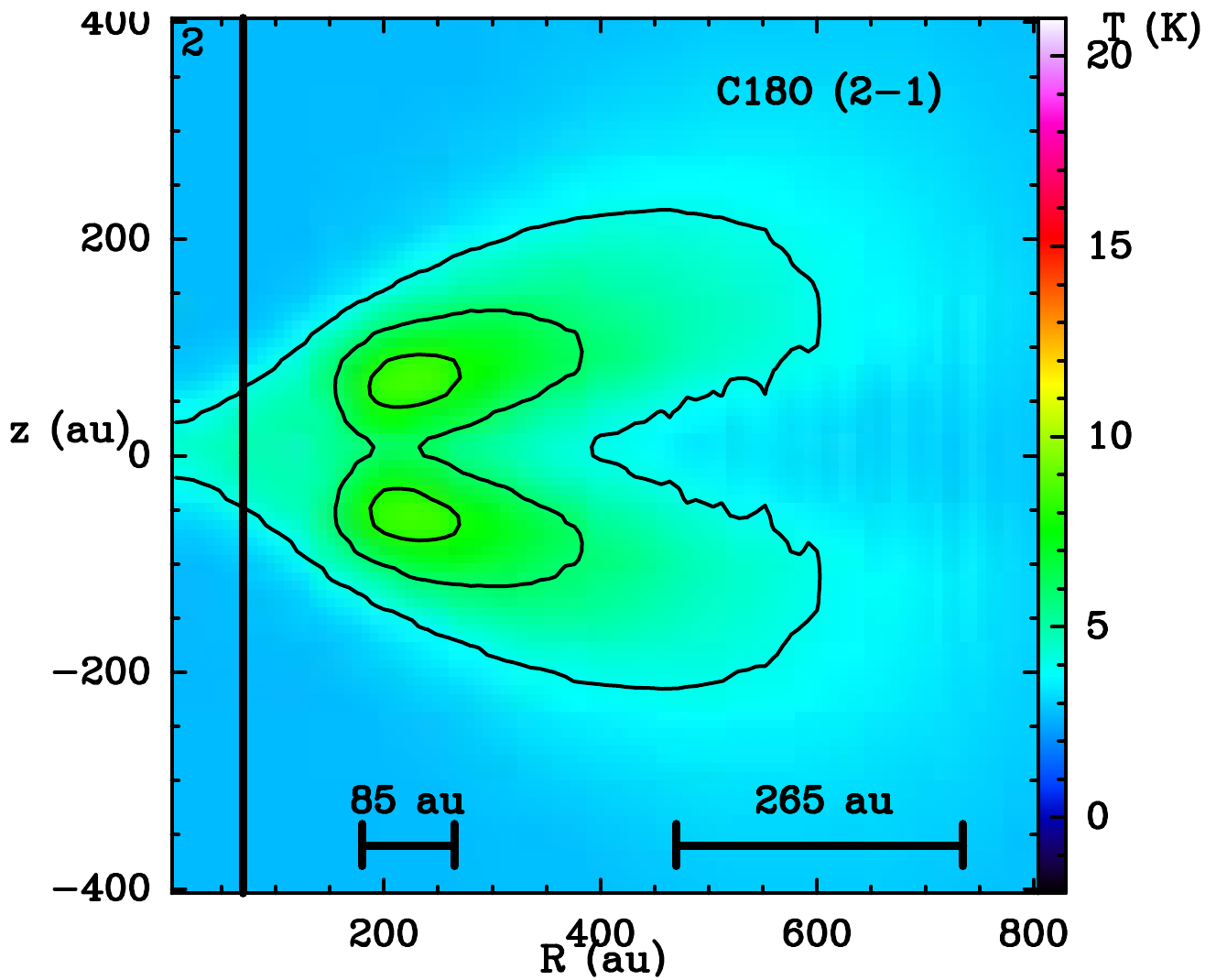}} 
  \subfigure{\includegraphics[width=0.268\textwidth]{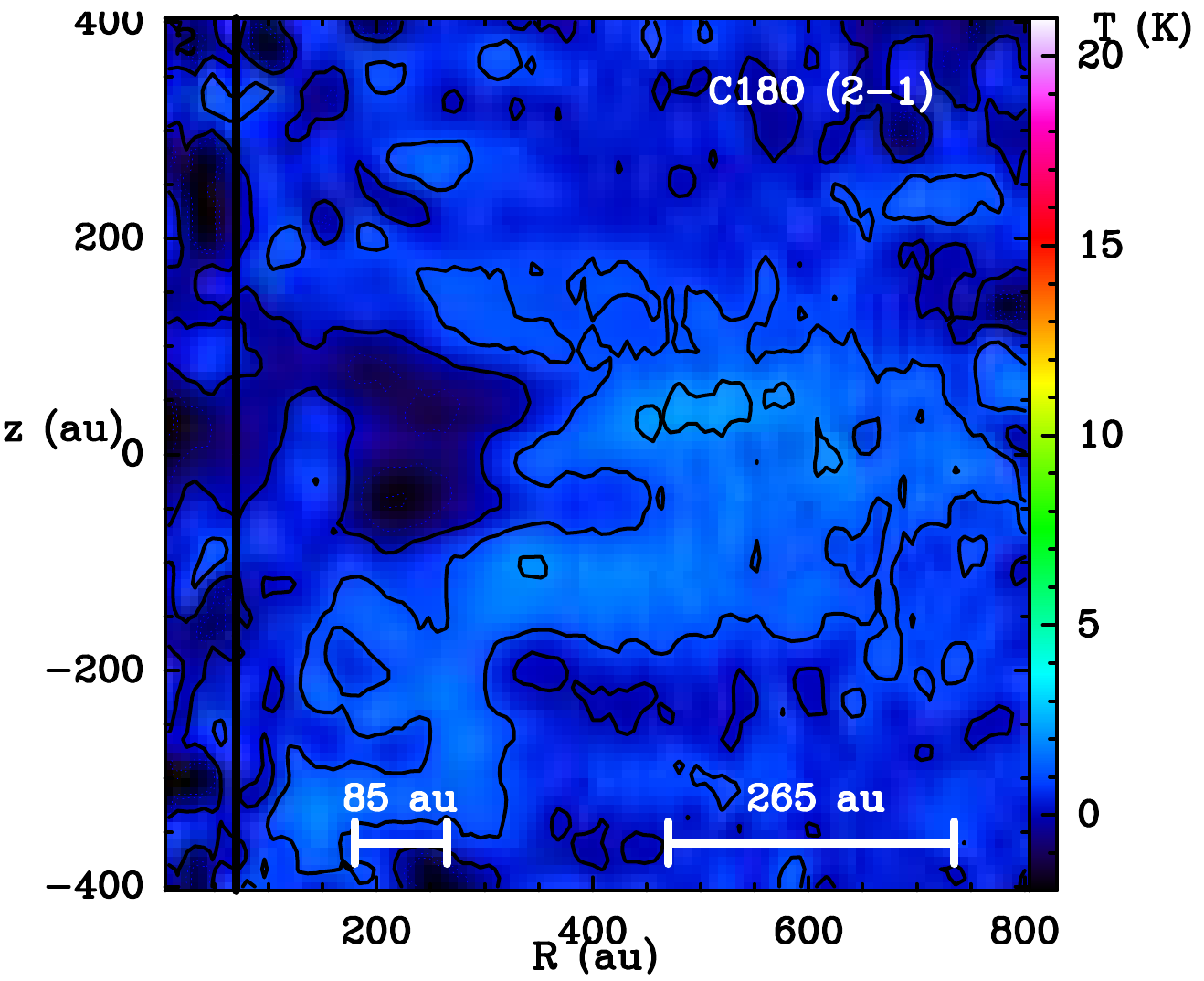}}

  \centering
  \subfigure{\includegraphics[width=0.268\textwidth]{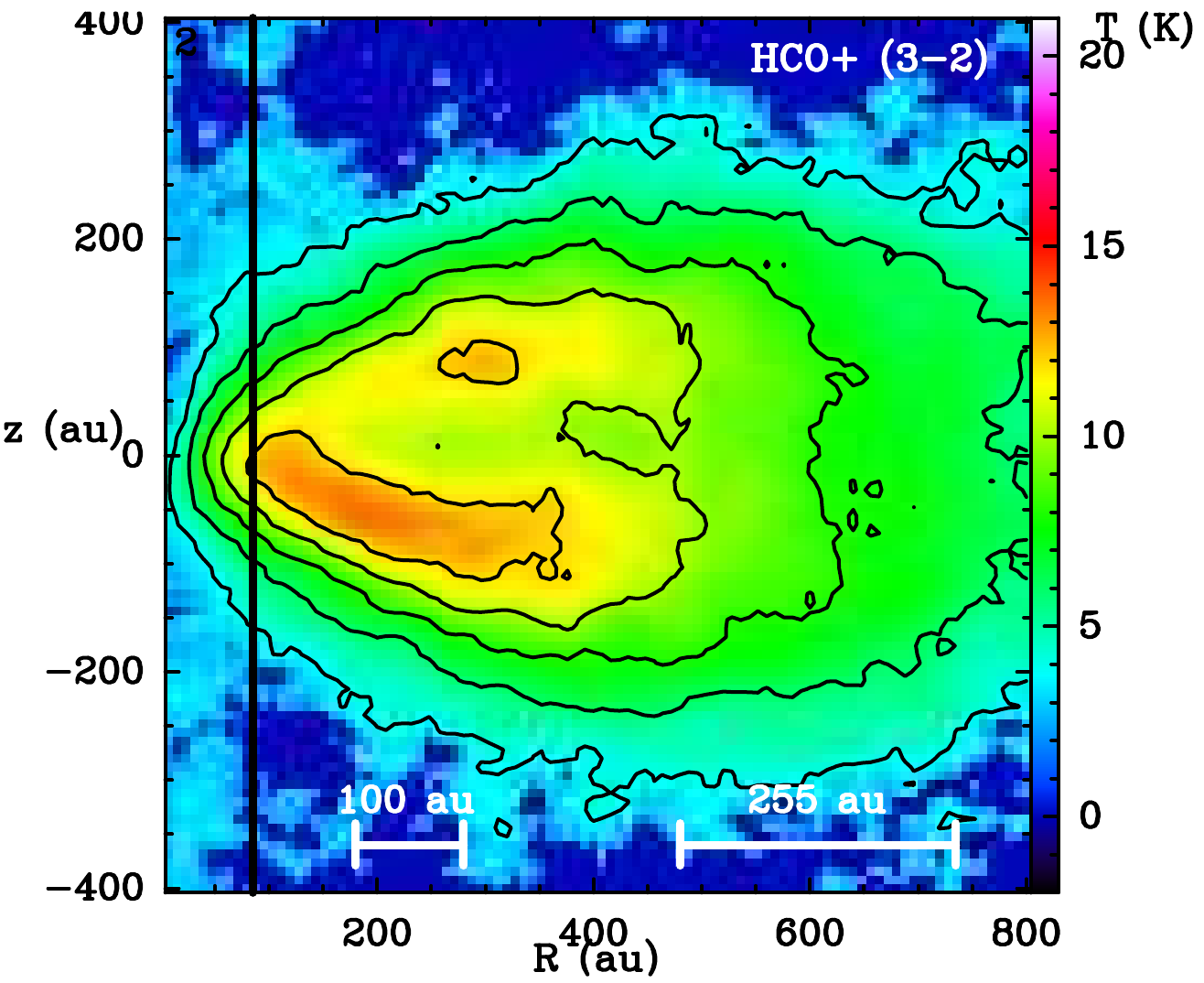}} 
  \subfigure{\includegraphics[width=0.268\textwidth]{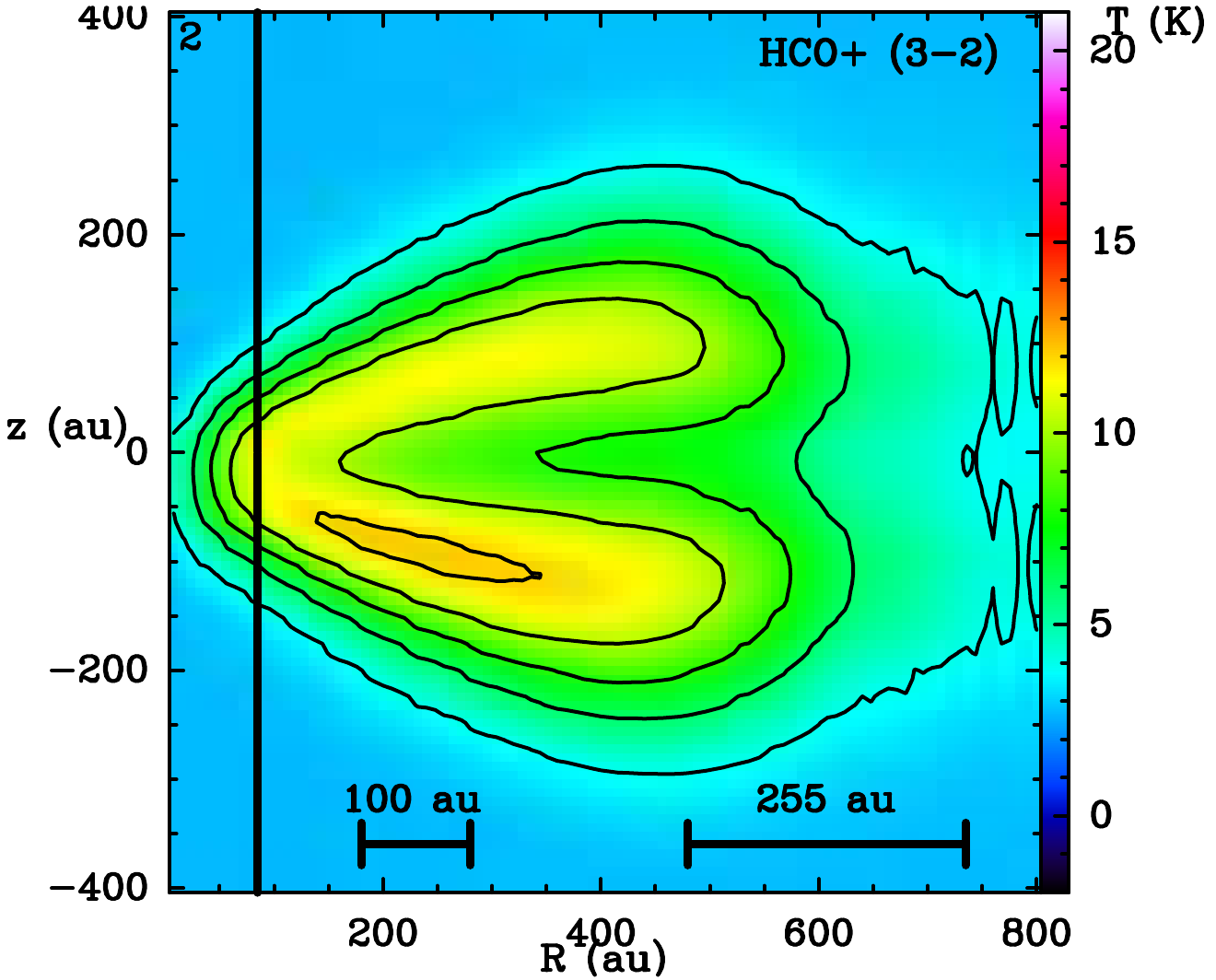}} 
  \subfigure{\includegraphics[width=0.268\textwidth]{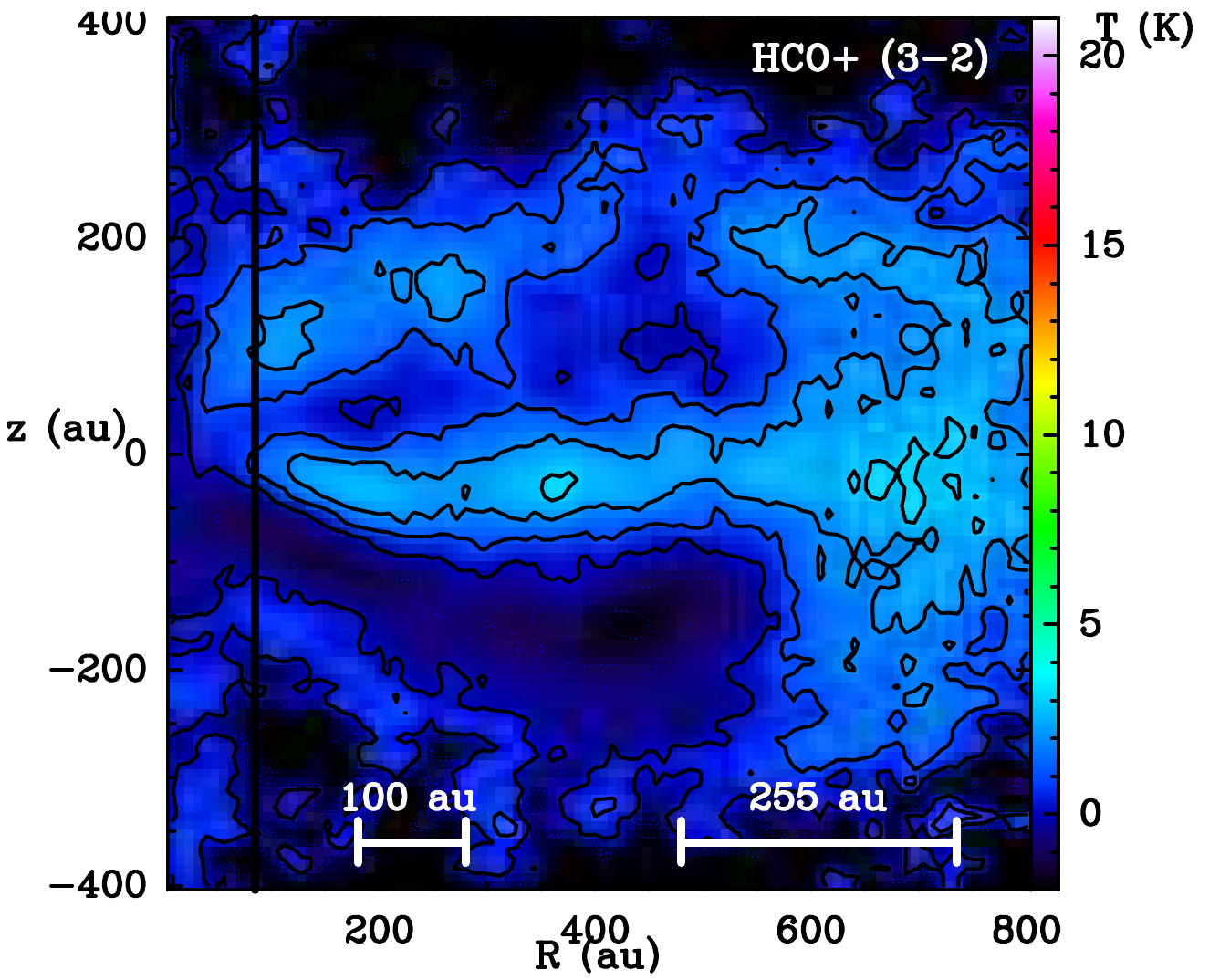}}
  
  \caption{From top to bottom: Tomographically Reconstructed Distribution (TRD) of $^{12}$CO 2-1, $^{12}$CO 3-2, $^{13}$CO 2-1, C$^{18}$O 2-1 and HCO$^+$ 3-2 : (a) TRD of the observations (b) TRD of the final result of the models (c) difference between the observations and models TRD. For $^{12}$CO (2–1), the residuals range from –2 to 2 K, while for the $^{12}$CO (3–2) transition, they extend from –3 to 2 K. In the case of $^{13}$CO (2–1), the differences are between 0 and 2 K. For C$^{18}$O (2–1), the deviations range from –1 to 2 K, and for HCO$^+$ (3–2), they span from –1 to 3 K. The vertical black line represents the effective resolution and two error bars show the widening of the beam as a function of radius caused by the TRD method. 
  The contours are defined from 4\,K to 20\,K in steps of 2\,K for observational and model maps, and from -3\,K to 3\,K in steps of 1\,K for the residual maps.}
 
\label{fig:tomo}
\end{figure*}
}

\begin{document}
   \title{Edge-On Disk Study (EODS) II: HCO$^+$ and CO vertical stratification in the disk surrounding SSTTau042021}
   \titlerunning{EODS II: HCO$^+$ and CO vertical stratification in the disk surrounding SSTTau042021}

   \author{C. Foucher  \inst{1}
          \and A. Dutrey \inst{1}   
          \and V. Pi\'etu \inst{2}
          \and S. Guilloteau \inst{1}
          \and E. Chapillon \inst{1,2}
         \and O.Denis-Alpizar \inst{3}
         \and E.Dartois \inst{4}
         \and E. Di Folco \inst{1}
         \and S. Gavino \inst{5}
         \and U. Gorti \inst{6}
         \and Th. Henning \inst{7}    
         \and \'A. K\'osp\'al \inst{7,8,9}
         \and F. Le Petit \inst{10}
         \and L. Majumdar \inst{11,12}
         \and R. Meshaka \inst{10}
          \and N. T. Phuong \inst{13}
          \and M. Ruaud \inst{6}
          \and D. Semenov \inst{7,14}
        \and Y.W. Tang \inst{15}
        \and S. Wolf \inst{16}
          }
    \authorrunning{C. Foucher  \inst{1}}
    
   \institute{Laboratoire d'Astrophysique de Bordeaux, Universit\'e de Bordeaux, CNRS, B18N, 
   All\'ee Geoffroy Saint-Hilaire, F-33615 Pessac\\
              \email{coralie.foucher@u-bordeaux.fr}
    \and IRAM, 300 Rue de la Piscine, F-38046 Saint Martin d'H\`{e}res, France
    \and  Departamento de F\'isica, Facultad de Ciencias, Universidad de Chile, Av. Las Palmeras 3425, \~Nu\~noa, Santiago, Chile
    \and Institut des Sciences Mol\'eculaires d'Orsay, CNRS, Univ. Paris-Saclay, Orsay, France 
    \and Dipartimento di Fisica e Astronomia, Universit\`a di Bologna, Via Gobetti 93/2, 40122, Bologna, Italy
    \and Carl Sagan Center, SETI Institute, Mountain View, CA, USA
    \and Max-Planck-Institut f\"{u}r Astronomie, K\"{o}nigstuhl 17, 69117 Heidelberg, Germany  
    \and Konkoly Observatory, HUN-REN Research Centre for Astronomy and Earth Sciences, MTA Centre of Excellence, Konkoly-Thege Mikl\'os \'ut 15-17, 1121 Budapest, Hungary
    \and Institute of Physics and Astronomy, ELTE E\"otv\"os Lor\'and University, P\'azm\'any P\'eter s\'et\'any 1/A, 1117 Budapest, Hungary
      \and LUX, Observatoire de Paris, PSL Research University,
CNRS, Sorbonne Universit\'es, 92190 Meudon, France
    \and Exoplanets and Planetary Formation Group, School of Earth and Planetary Sciences, National Institute of Science Education and Research, Jatni 752050, Odisha, India
    \and Homi Bhabha National Institute, Training School Complex, Anushaktinagar, Mumbai 400094, India
    \and Department of Astrophysics, Vietnam National Space Center, Vietnam Academy of Science and Technology, 18 Hoang Quoc Viet, Cau Giay, Hanoi, Vietnam
     \and Zentrum f\"{u}r Astronomie der Universit\"{a}t Heidelberg, Institut f\"{u}r Theoretische Astrophysik, Albert-Ueberle-Str. 2, 69120 Heidelberg, Germany  
    \and Academia Sinica Institute of Astronomy and Astrophysics, 11F of AS/NTU Astronomy-Mathematics Building, No.1, Sec. 4, Roosevelt Rd, Taipei 106319, Taiwan, R.O.C
    \and Institut f\"ur Theoretische Physik und Astrophysik, Christian-Albrechts-Universit\"at zu Kiel, Leibnizstrasse 15, 24118 Kiel, Germany}

   \date{ }

  \abstract
  {Edge-on disks offer a unique opportunity to directly examine their vertical structure, providing valuable insights into planet formation processes. We investigate the dust properties, as well as the CO and HCO$^+$ gas properties, in the edge-on disk surrounding the T Tauri star 2MASS J04202144+281349 (SSTTau042021).}
    {We estimate the radial and vertical temperature and density profile for the gas and the dust.}
   {We use ALMA archival data of CO isotopologues and continuum emission at 2, 1.3 and 0.9 mm together with new NOEMA HCO$^+$ 3-2 observations. We retrieve the gas and dust disk properties using the tomographic method and the \textsc{DiskFit} model.}
  {The vertical CO emission appears very extended, partly tracing the H$_2$ wind observed by JWST. C$^{18}$O, $^{13}$CO and HCO$^+$ emission characterize the bulk of the molecular layer. The dust and gas have a mid-plane temperatures of $\sim 7-11$ K. The temperature of the molecular layer (derived from $^{13}$CO and HCO$^+$) is on the order of 16 K. HCO$^+$ 3-2 being thermalized, we derive a lower limit for the H$_2$ volume density of $\sim 3 \times 10^6$ cm$^{-3}$ at radius 100-200 au between 1 and 2 scale heights. The atmosphere temperature of the CO gas is of the order $\sim$ 31 K at a radius of 100 au. We directly observe CO and HCO$^+$ gas onto the mid-plane beyond the dust outer radius ($\ge 300$ au). The (gas+dust) disk mass estimated up to a radius of 300 au is on the order of $4.6\times10^{-2} \Msun$.}
   {Thanks to the favorable disk inclination, we present the first quantitative evidence for vertical molecular stratification with direct observation of CO and HCO$^+$ gas along the mid-plane. We estimate the temperature profile with temperature of 7-11 K near the mid-plane, and 15-20 K in the dense part of the molecular layer up to $\sim$ 35 K above.}


   \maketitle

\section{Introduction}

Gas-rich protoplanetary disks are natural laboratories to study planet formation mechanisms. 

Nowadays, large mm/submm arrays (ALMA and NOEMA) enable us to study their gas and dust properties with sufficient angular resolution and sensitivity to start investigations on planet forming regions. For instance, the ALMA large program Disk Substructures at High Angular Resolution Project (DSHARP) observed rings and spirals in dust disks that may be due to active planet formation \citep{Andrews+2018}. 

The ALMA large program MAPS (Molecules with ALMA at Planet forming Scales) provided a significant contribution to the understanding of the gas phase by observing at high-resolution five large inclined disks  around TTauri and Herbig Ae stars. It delivered unprecedented insights into the gas temperature, density and composition. Key molecules such as HCO$^+$, CN, HCN, HNC, CS and N$_2$H$^+$ were observed \citep{Guzman+2021}. Among them, HCO$^+$, plays a crucial role because it provides invaluable insights into the ionization structure of disks, especially in regions where other tracers become less reliable \citep{Aikawa+2021}. Models suggest that HCO$^+$ should be emitted from a region similar to CO, closer to the disk mid-plane \citep{Aikawa+2002}.

Accurate CO observations allow the retrieval of the gas kinetic temperature since the first levels of its rotational lines are thermalized and optically thick \citep[]{Dartois+etal_2003,Franceschi+2024}. In five disks, a parametric approach has been used to derive the radial and vertical temperature profiles \citep{Law+2021} and a more complex one using a thermo-chemical model \citep{Zhang+etal_2021}. The comparison shows that both approaches provide similar thermal structures in the 15-40 K temperature range \citep{Zhang+etal_2021}. There were recently several attempts to improve  these methods \citep[e.g.][]{Paneque+2023}.

These retrieval methods require to identify an emitting surface or thin layer, but their performance would degrade for geometrically thick layers probed by optically thin transition because of the combined effects of density, temperature and velocity gradients along the line
of sight that make the identification of the sampled  $(r,z)$ locations difficult.

Edge-on disks are not affected by this limitation, as the disk radii appear as a straight line in the velocity-position diagram that provides a Tomographically Reconstructed Distribution (TRD) of the disk structure by averaging the molecular emission along each radius \citep{Dutrey+etal_2017}. 

If the observed line is optically thick and thermalized, such as the $^{12}$CO lines, the gas temperature can be directly derived \citep{Dutrey+etal_2017,Pinte+2018,Flores+etal_2021}. For an optically thinner molecular transition, the vertical distribution of the observed molecule is directly estimated. 

Finally, ALMA observations of dust disks have also offered accurate evidence of dust settling at millimeter wavelengths \citep[e.g.][]{Guilloteau+2016,Villenave+etal_2020,Villenave+etal_2022,Franceschi+2023}, but the molecular stratification of the disk remains largely unexplored. 
\begin{figure*}[!h]
  \centering
  
  \subfigure(a){\hspace{5cm}} 
  \subfigure(b){\hspace{5cm}} 
  \subfigure(c){}\\
  
  \centering
  \subfigure{\includegraphics[width=0.3\textwidth]{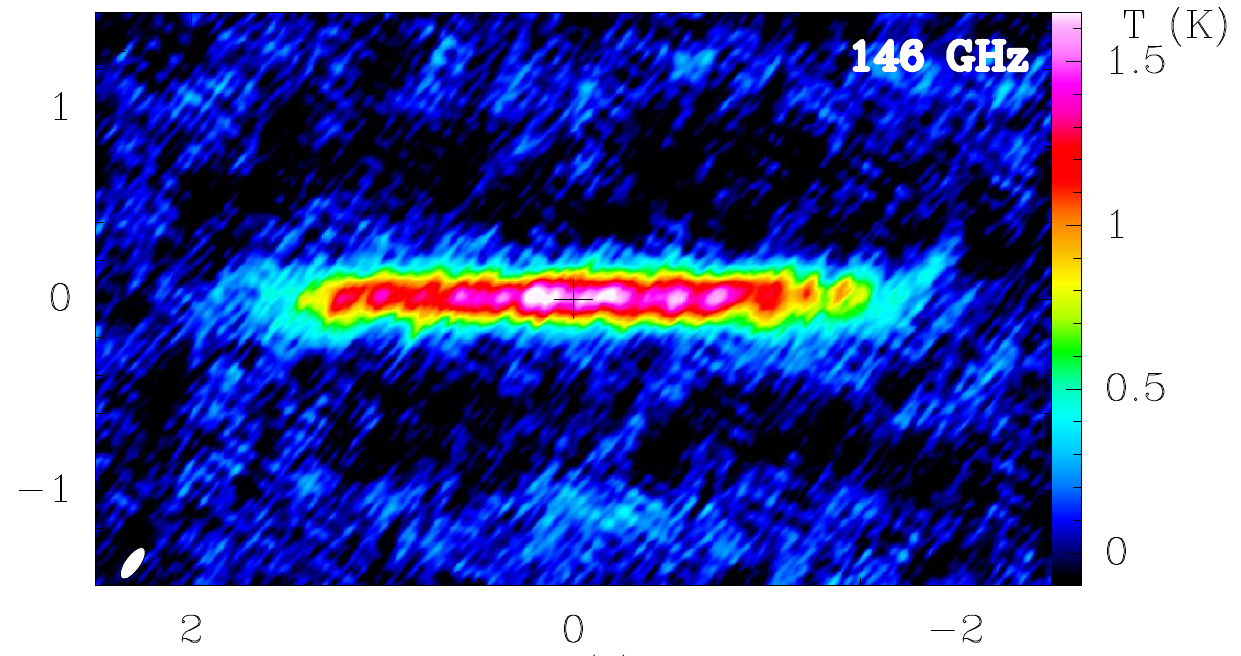}} 
  \subfigure{\includegraphics[width=0.3\textwidth]{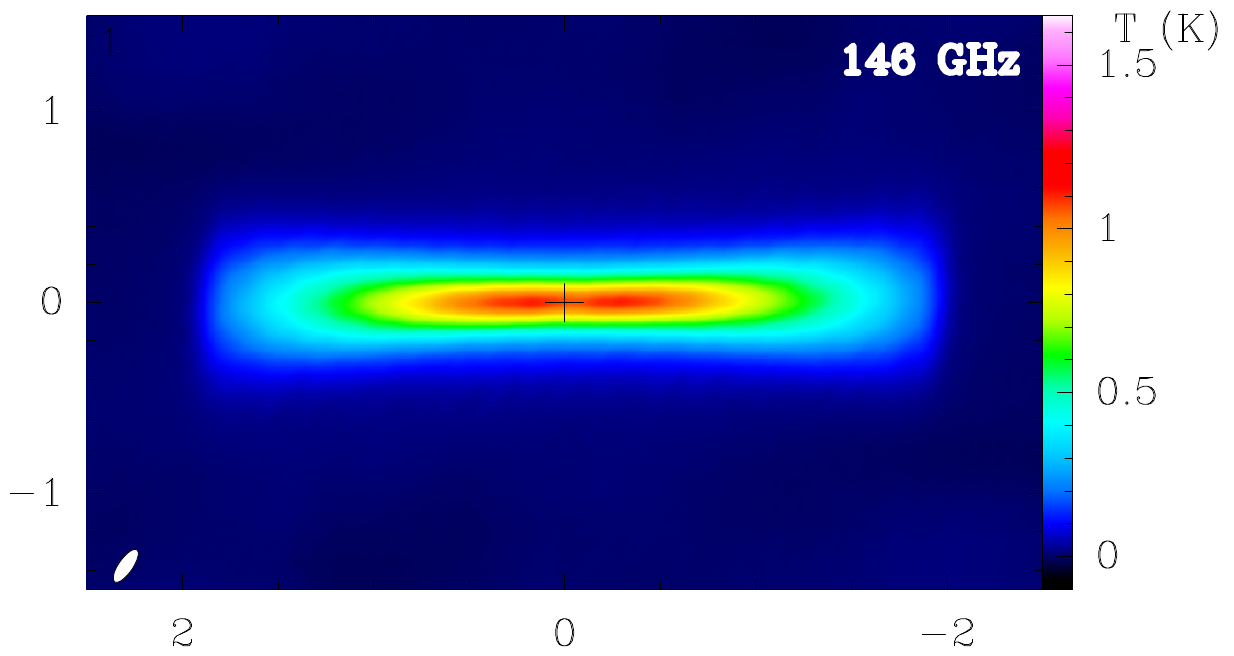}} 
  \subfigure{\includegraphics[width=0.3\textwidth]{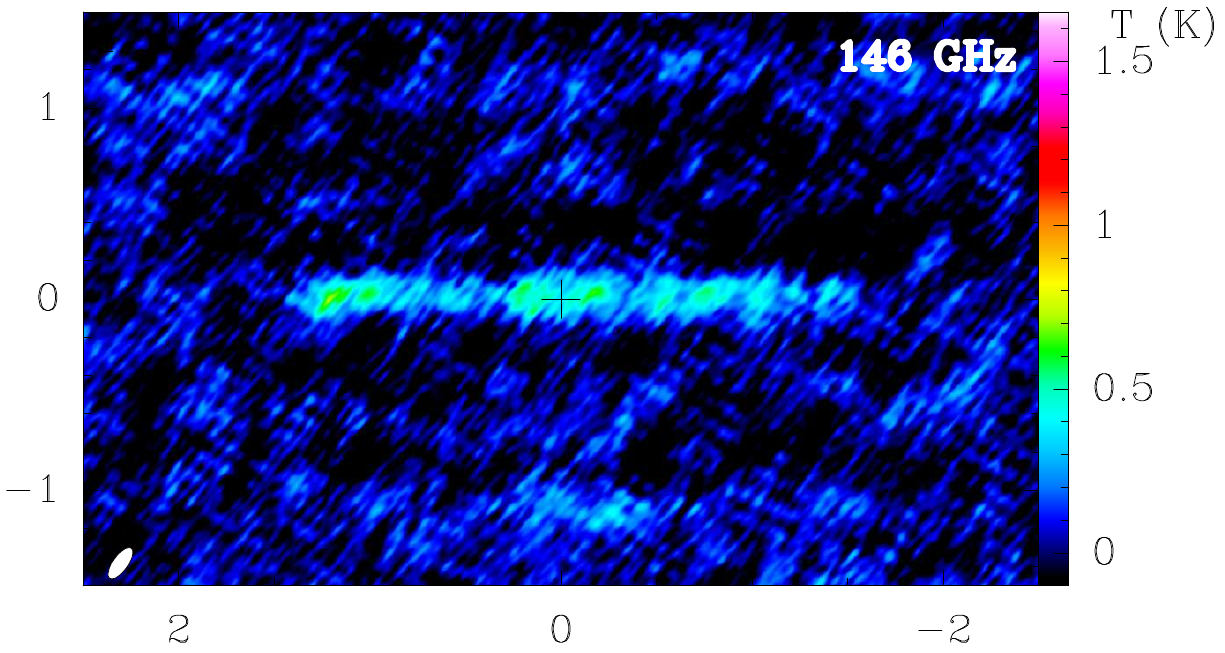}}

  \subfigure{\includegraphics[width=0.3\textwidth]{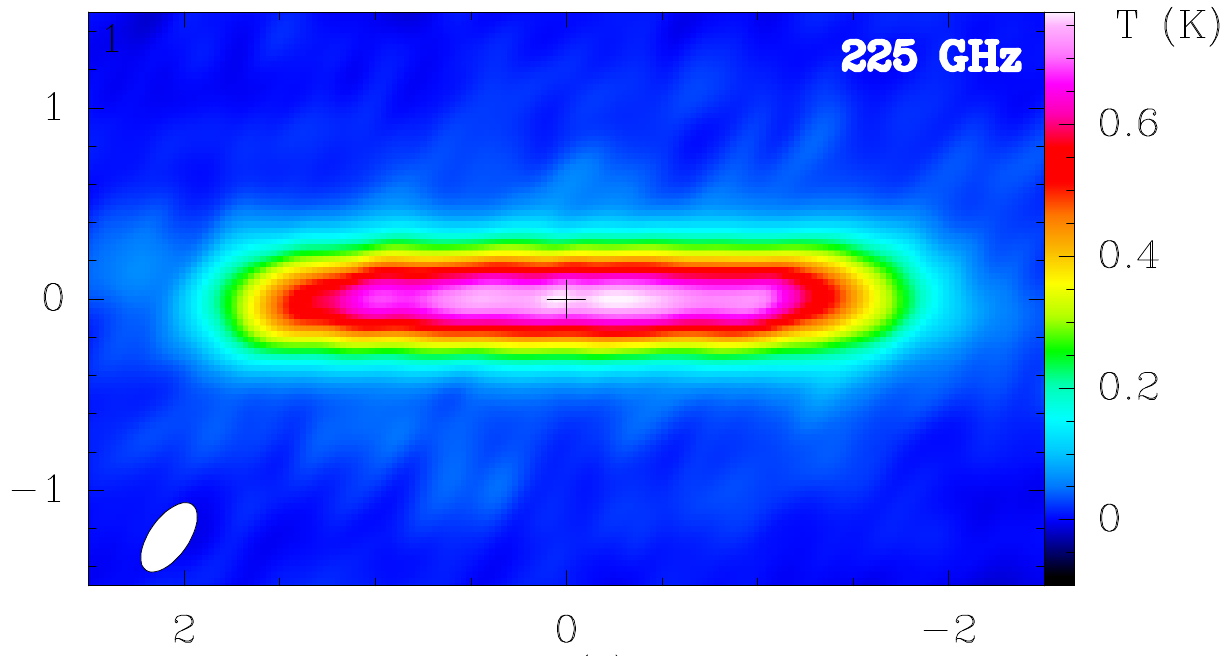}} 
  \subfigure{\includegraphics[width=0.3\textwidth]{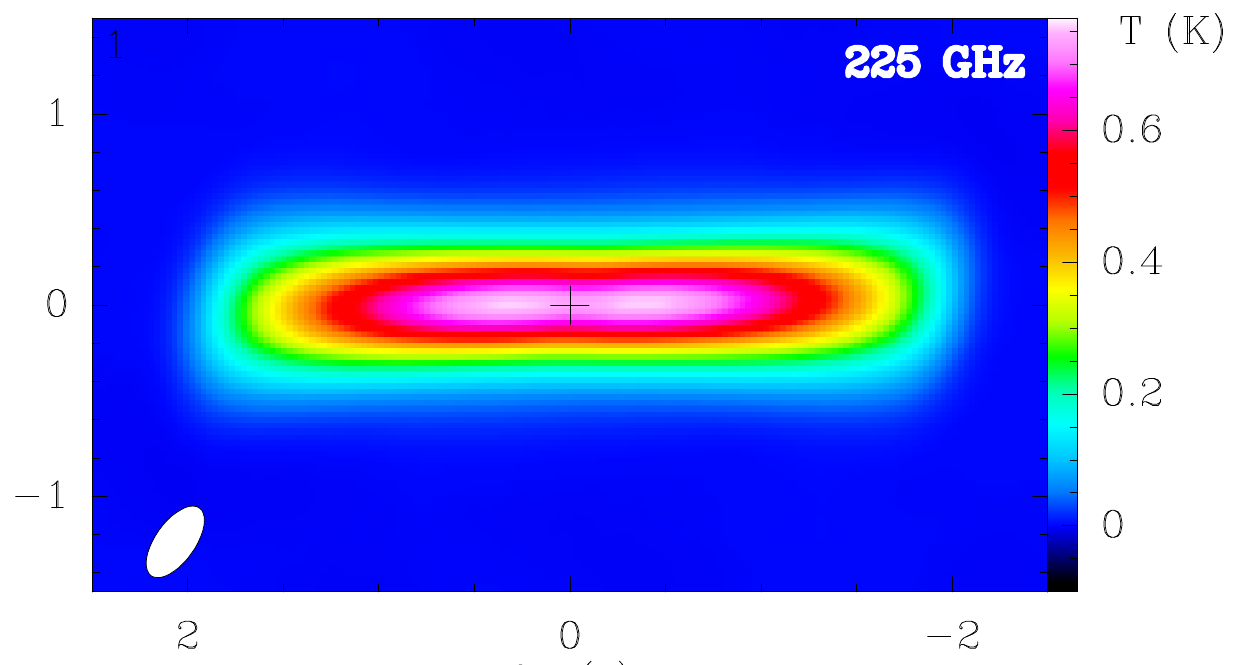}} 
  \subfigure{\includegraphics[width=0.3\textwidth]{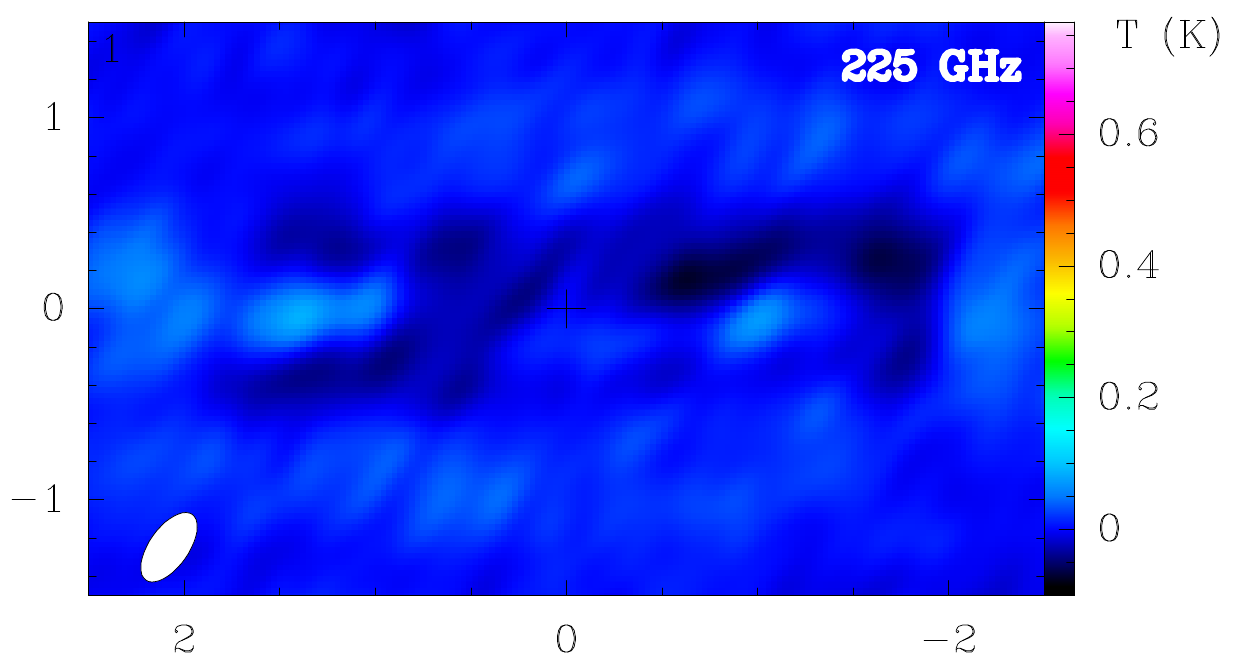}}

  \subfigure{\includegraphics[width=0.3\textwidth]{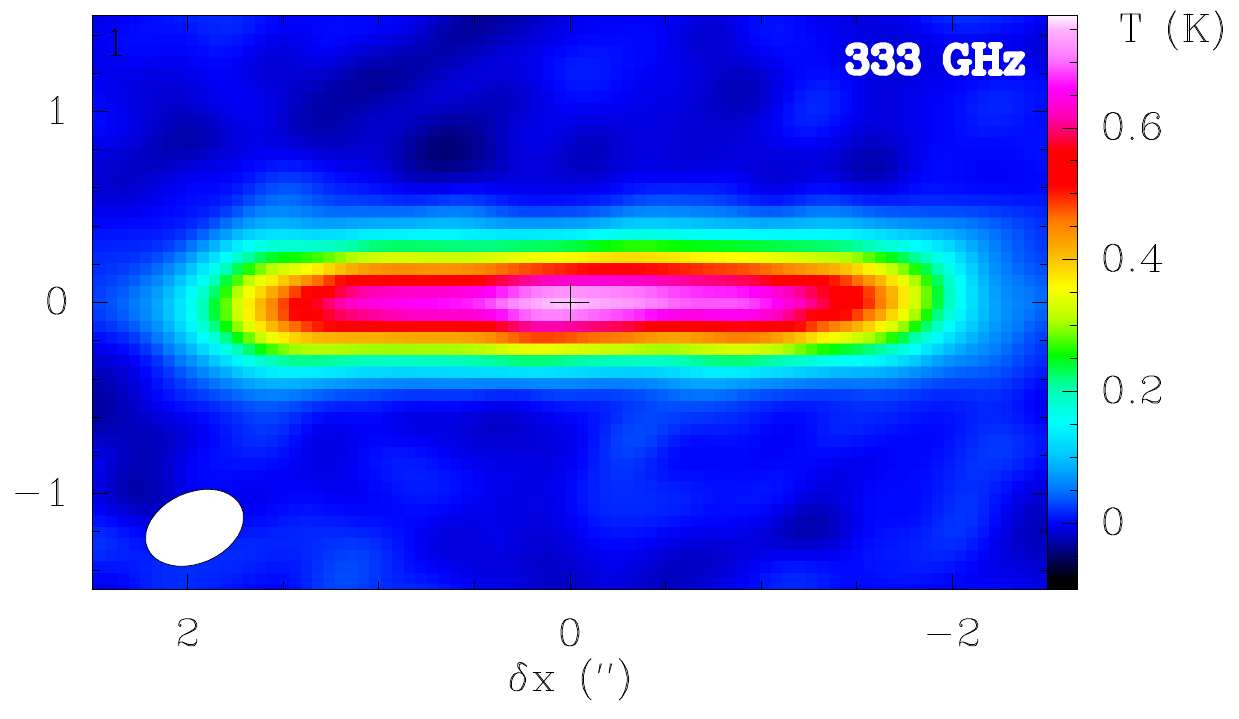}} 
  \subfigure{\includegraphics[width=0.3\textwidth]{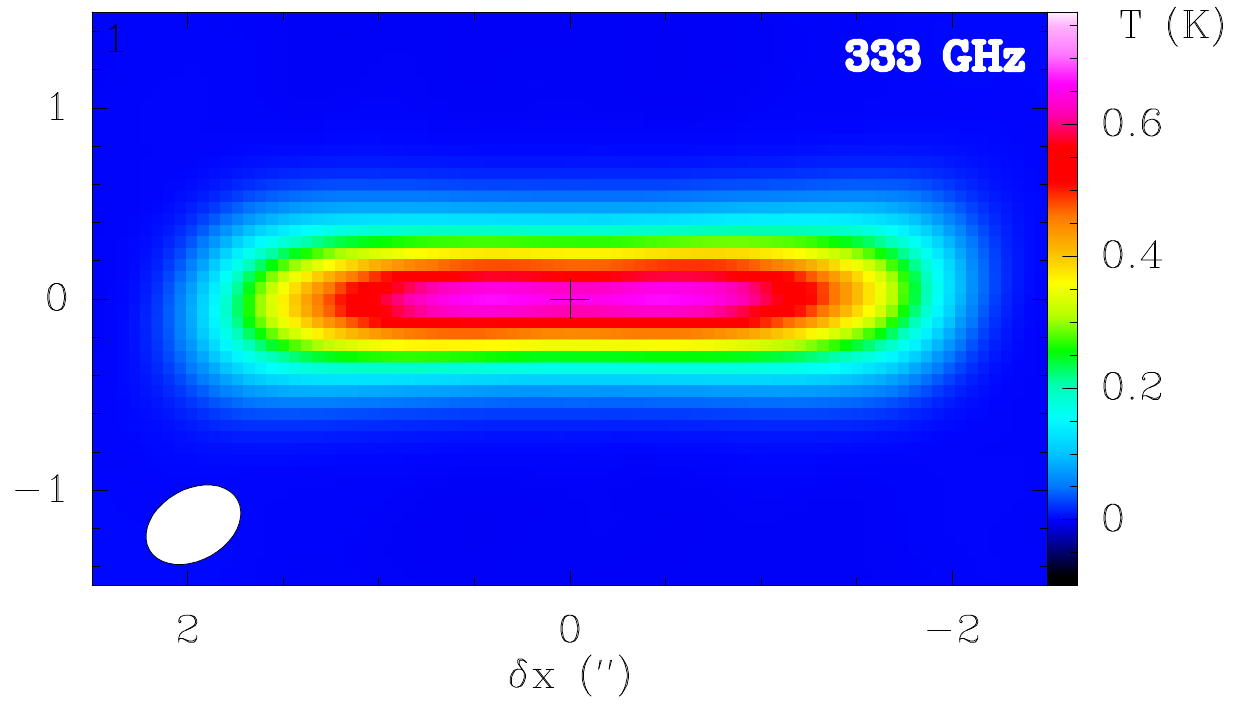}} 
  \subfigure{\includegraphics[width=0.3\textwidth]{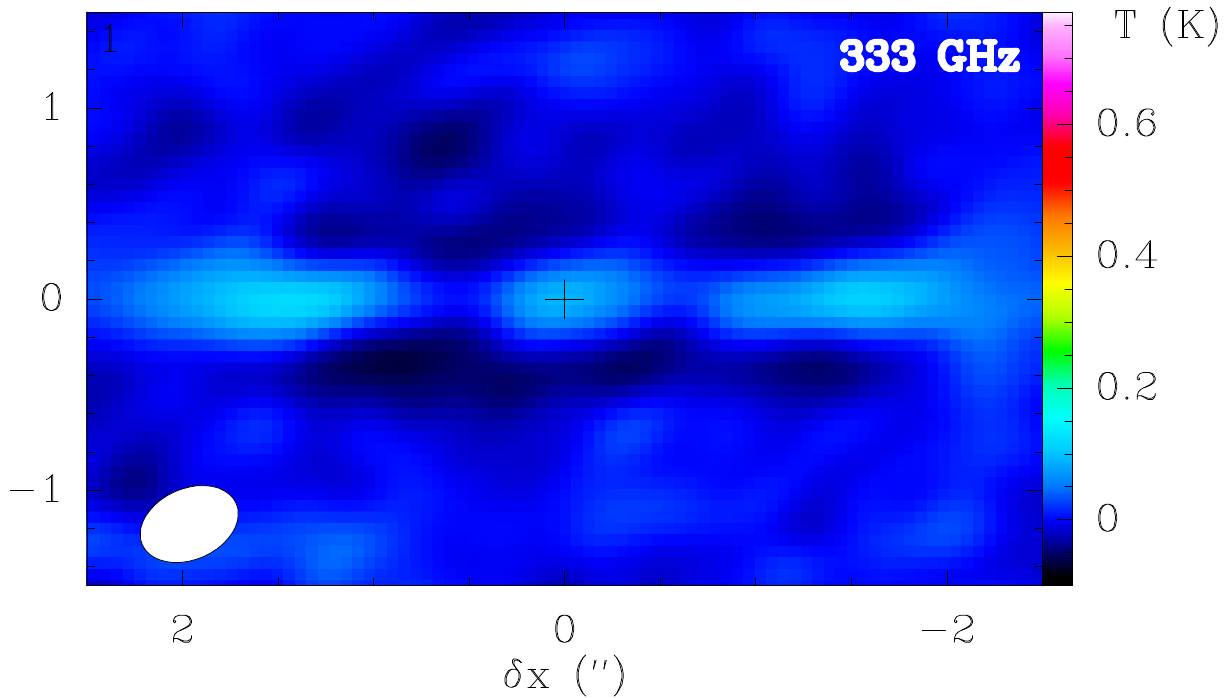}}

  \caption{Continuum observations: Left column (a): observation, middle column (b): best model, right column (c): difference (observation - model). The rms noise levels reach $0.1 \mathrm{K}$ at 146 GHz, $6.55 \times 10^{-3}\mathrm{K}$ at 225 GHz, and $1.33 \times 10^{-2}\mathrm{K}$ at 333 GHz. The peaks of the residuals correspond to $5.2\sigma$ at 146 GHz, $11.8\sigma$ at 225 GHz, and $8.7\sigma$ at 333 GHz.} 
  \label{fig:dust}
\end{figure*}
In this paper, we analyze new NOEMA observations of HCO$^+$ 3-2 together with ALMA Archival CO and continuum data of the disk orbiting around 2MASS J04202144+2813491 (hereafter SSTTau042021). This is a Class II TTauri star of mass 0.27\,M$_{\odot}$ \citep{Simon+2019} located in the Taurus cloud at a distance of $\sim160$ pc. The dust disk is very close to edge-on (inclination $>$ 85 $^{\circ}$), is settled and has an outer radius of $\sim350$ au, while the CO disk extends up to $\sim700$ au \citep{Villenave+etal_2020}. More recently, JWST observations by \citet{Arulanantham+2024} and \citet{Duchene+2024} revealed extended vertical emission from H$_2$ rotational transitions with an X shape, likely due to an MHD disk wind extending up to about 200 au above the mid-plane (see also Fig. \ref{fig:jwst}).

\begin{table*}
\caption{ALMA and NOEMA continuum and line observations.} 
\centering
\begin{tabular}{c|c|cc|c|c|c|c}
\hline

 Line   & Spectral & Angular  & P.A. & Peak flux & Sensitivities & Total & Maximum Recoverable \\
        & Resol. & Resol. & & densities  &  per channel & Flux & Scale (MRS) \\
        & (km/s) & ($'' \times ''$) & ($^\circ$) & (mJy/beam) & (mJy/beam) & (Jy.km/s) & (")\\
 \hline
 $^{12}$CO 3-2   & 0.21  &   $0.52 \times 0.36$ & 112 & 335 & 16.3 & 6.92 $\pm$ 0.24 & 7.2 \\ 
 $^{12}$CO 2-1   & 0.16 &   $0.41 \times 0.2$ & 144 & 67.2 & 2.98 & 6.22 $\pm$ 0.13 & 10.7 \\
$^{13}$CO 2-1   & 0.08  &   $0.43 \times 0.21$ & 144 & 51.1 & 2.98 & 4.59 $\pm$ 0.11 & 11.2 \\
C$^{18}$O 2-1   & 0.08 &   $0.43 \times 0.21$ &  144 & 22.3 & 2.25 & 1.34 $\pm$ 0.27 & 11.3 \\
  HCO$^+$ 3-2   & 0.07 &   $0.56 \times 0.30$ &  122 & 117 & 8.69 & 4.94 $\pm$ 0.17 & 6.1 \\
\hline 
\hline
Continuum  & Frequencies & Angular  & P.A. & Peak flux & Sensitivities &  Total & Maximum Recoverable \\
        & & Resol. & & densities  & & Flux & Scale (MRS)\\
  & (GHz) & ($'' \times ''$) & ($^\circ$)& (mJy/beam) & (mJy/beam) & (mJy) & (") \\ 
\hline  
 & 146 & $0.10 \times 0.04$ & 30 & 0.22 & 0.02 & 16.9 & 16.5\\
 & 225 & $0.42 \times 0.21$ & 145 & 2.75 & 0.01 & 44 & 10.7 \\
 & 333 & $0.54 \times 0.37$\ & 114 & 13.67 & 0.13 & 95 & 7.2 \\
\hline
\hline
\end{tabular}
\tablefoot{The total flux was measured by integrating the emission over the entire disk, without thresholding, using the integration tool available in the IMAGER software. For the line emission, the total flux was computed over specific velocity ranges: from 4.35 to 10.83 km/s for $^{12}$CO (2–1), 4.64 to 10.53 km/s for $^{12}$CO (3–2), 5.37 to 9.11 km/s for $^{13}$CO (2–1), 5.17 to 9.05 km/s for C$^{18}$O (2–1), and 4.95 to 9.48 km/s for HCO$^+$ (3–2). The MRS is given by $0.6 \times \lambda / B_{min}$, with $B_{min}$ = 15 meters for ALMA and $B_{min}$ = 16 meters for NOEMA.}
\label{tab:obs}
\end{table*}

\begin{figure*}[!ht]
  \centering
  \includegraphics[width=8cm]{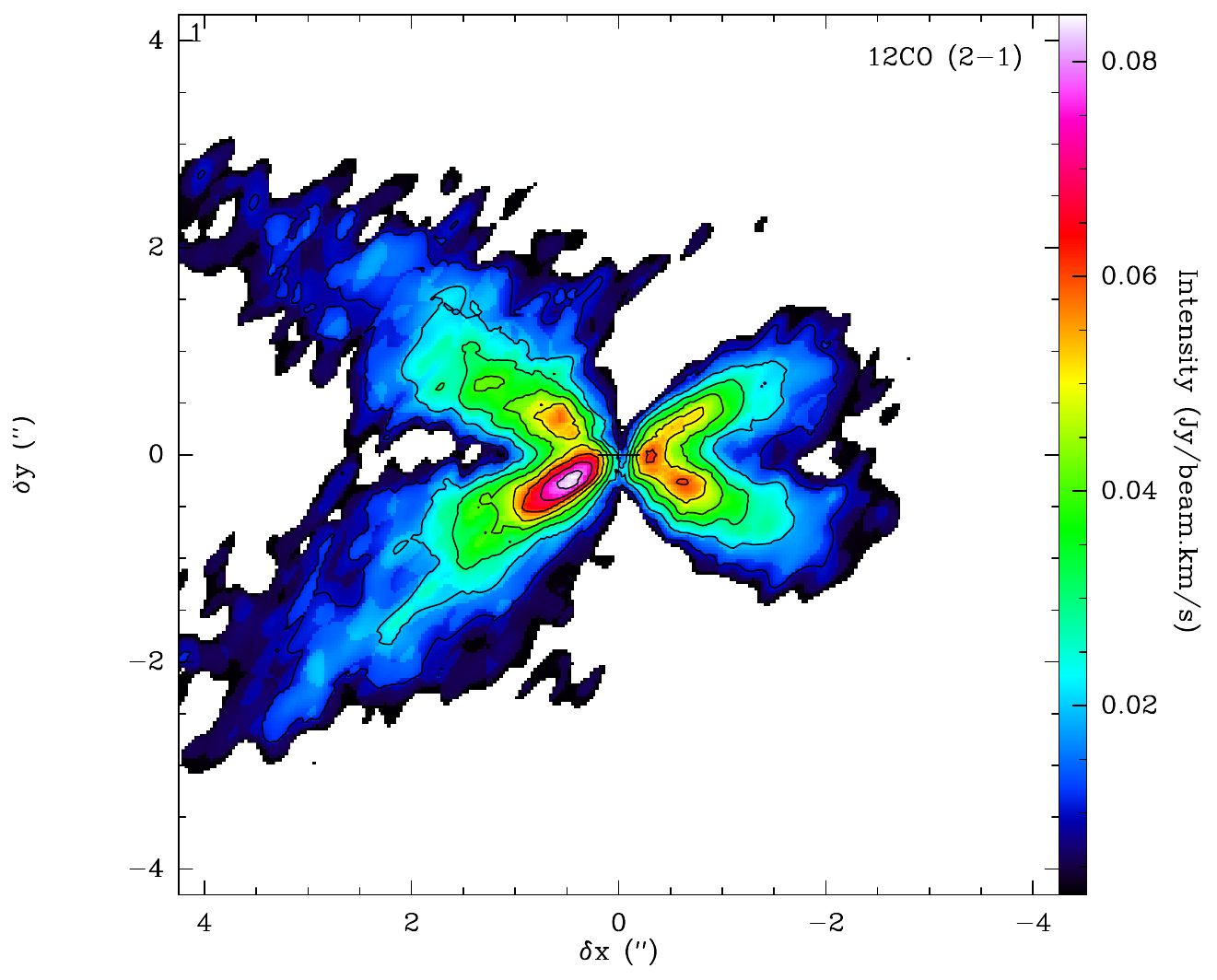}
  \includegraphics[width=8cm]{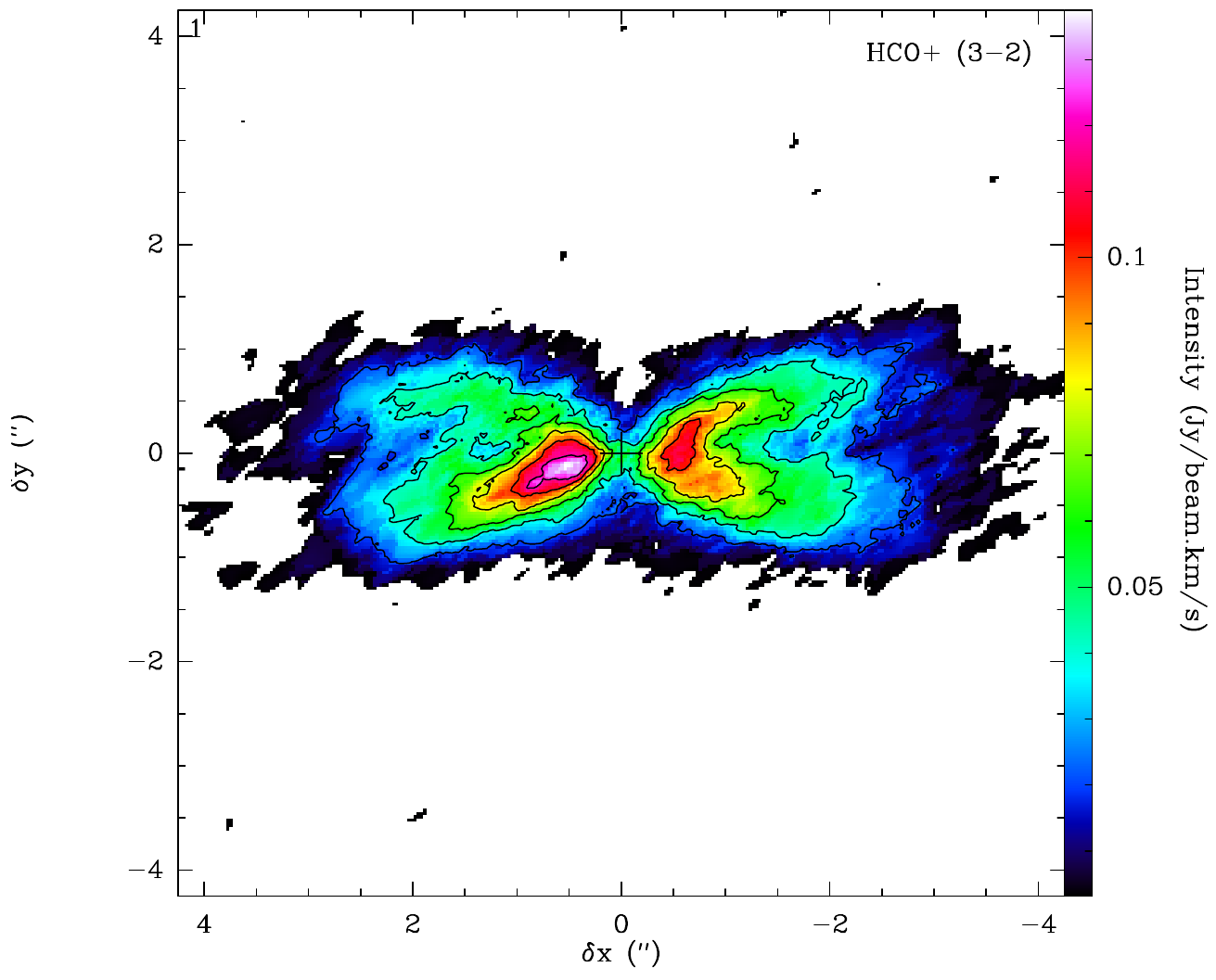}
  \caption{On the left is an integrated intensity map of $^{12}$CO 2-1 (clipped at 22 mJy/beam.km/s), and on the right is an integrated intensity map of HCO$^{+}$ 3-2 (clipped at 8 mJy/beam.km/s). The maps are rotated by -73.5$^\circ$. The contours are defined from 0 to 0.1 Jy/beam.km/s in steps of 0.01 Jy/beam.km/s, i.e 3.36$\sigma$ for the $^{12}$CO 2-1 map, and from 0 to 0.5 Jy/beam.km/s in steps of 0.02 Jy/beam.km/s , i.e 2.30$\sigma$ for the HCO$^{+}$ 3-2 map.}
    \label{fig:moments}
\end{figure*}

\section{Observations and Data Reduction}
\subsection{Observations}
We used ALMA archival data for CO isotopologues and continuum at related frequencies, as well as for the 145 GHz continuum, and observed the HCO$^+$ 3-2 line with the NOEMA interferometer.

Table \ref{tab:obs} presents the main observational characteristics of the data.

\paragraph{NOEMA:}
The new long baselines (up to 1600 m) were used to observe SSTTau042021 in HCO$^+$ 3-2 and in continuum in winter 2022/2023 (project W22BC). 
These data were completed by observations in C configuration (baselines range 16 to 350\,m) in Nov.\,2024, project S24AT) to provide more sensitivity to extended emission.
Data were calibrated using the standard CLIC pipeline. 

\paragraph{ALMA:}
Observations were performed in projects 2016.1.00771.S for the $^{12}$CO, $^{13}$CO and C$^{18}$O 2-1 lines and the continuum at 230 GHz and 2016.1.00460.S for the $^{12}$CO 3-2 transition and the continuum at 345 GHz and 150 GHz (Table \ref{tab:obs}). Project 2016.1.00771.S was observed in configuration C40-6, with baselines ranging from 15 to 1800 meters, and project 2016.1.00460.S was observed in configuration C40-4, with baselines from 15 to 704 meters. Data were calibrated with the standard ALMA pipeline \citep[see][for details of the CO 
line data]{Simon+2019} and exported in UVFITS format for further use.

\begin{figure*}[!ht]
\begin{frame}
\centering
  \includegraphics[width=5.75cm]{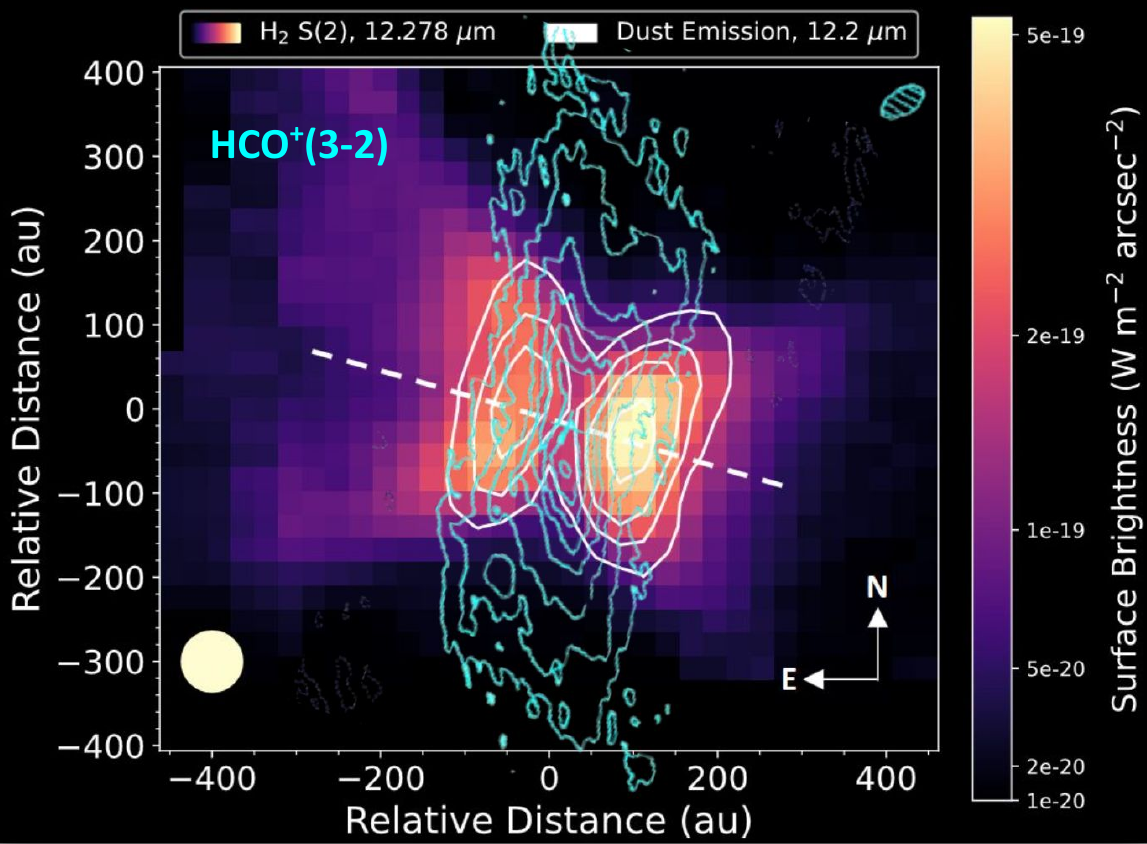}
  \centering
  \includegraphics[width=5.75cm]{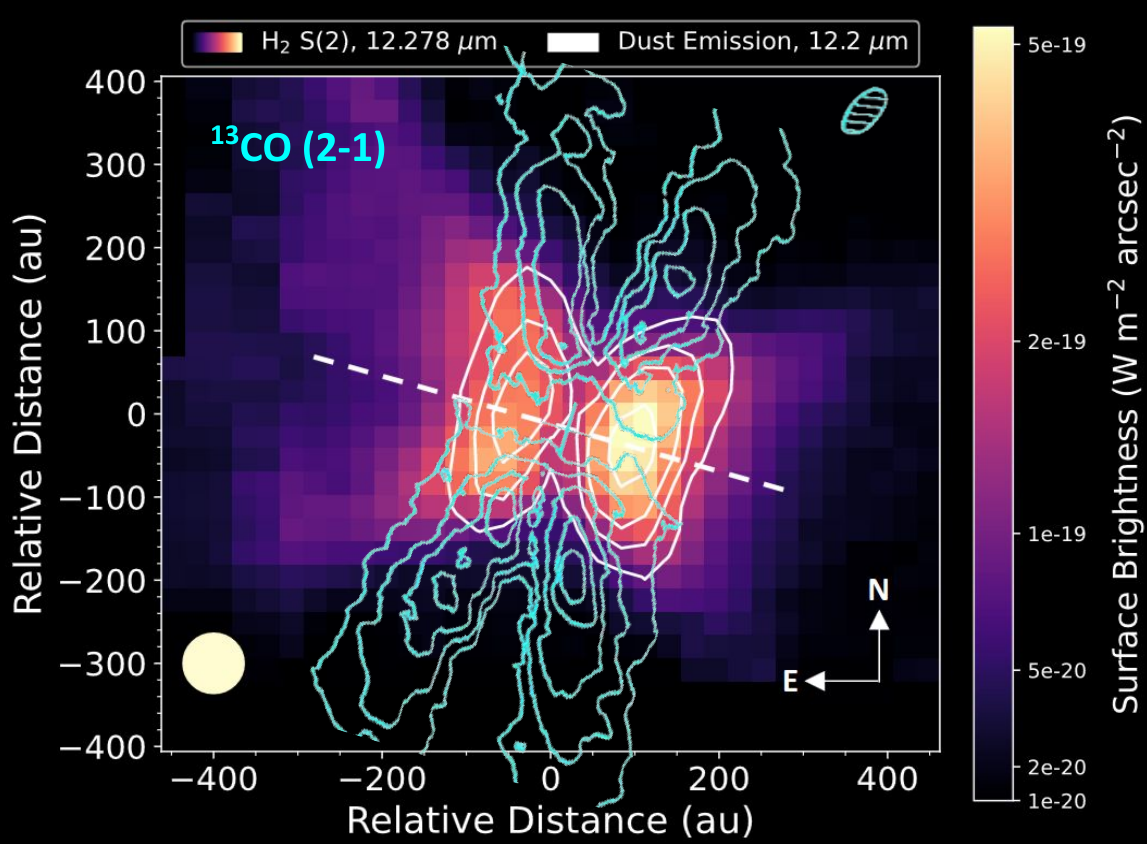}
  \centering
  \includegraphics[width=5.90cm]{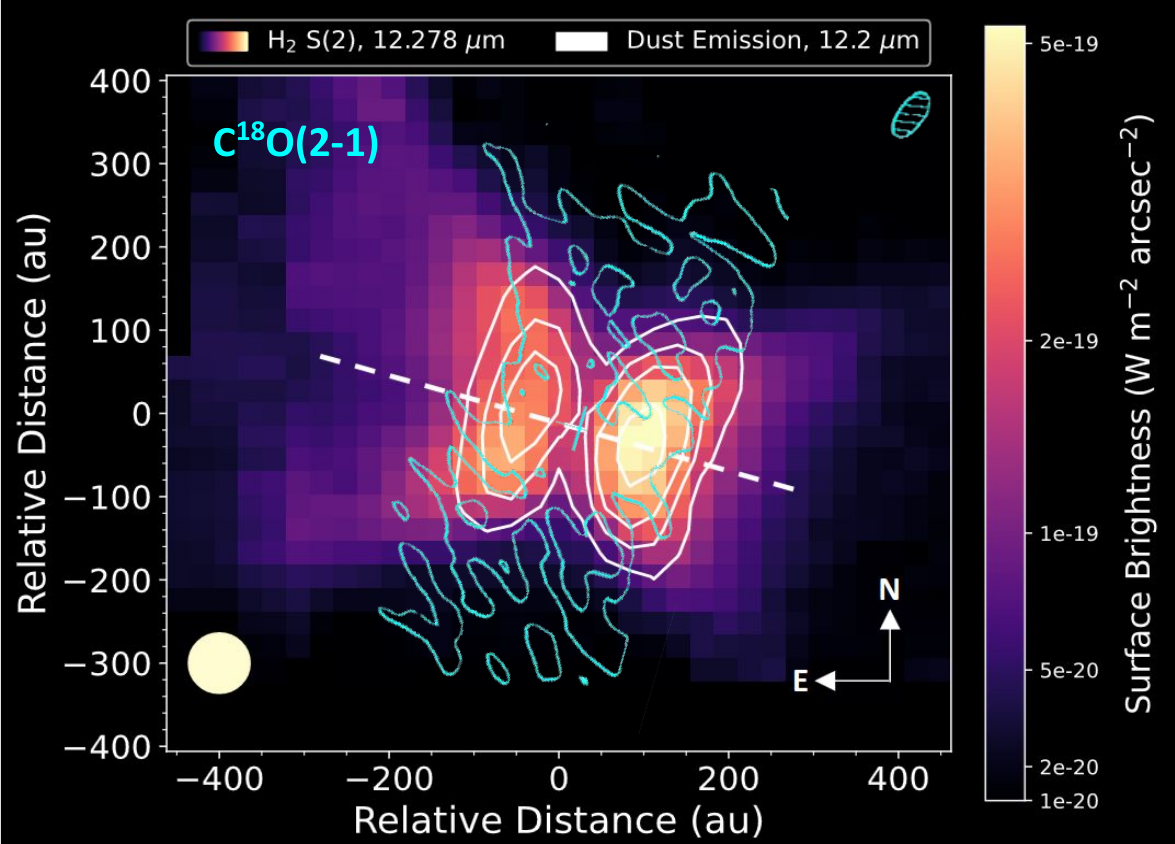}
  \centering
  \includegraphics[width=5.75cm]{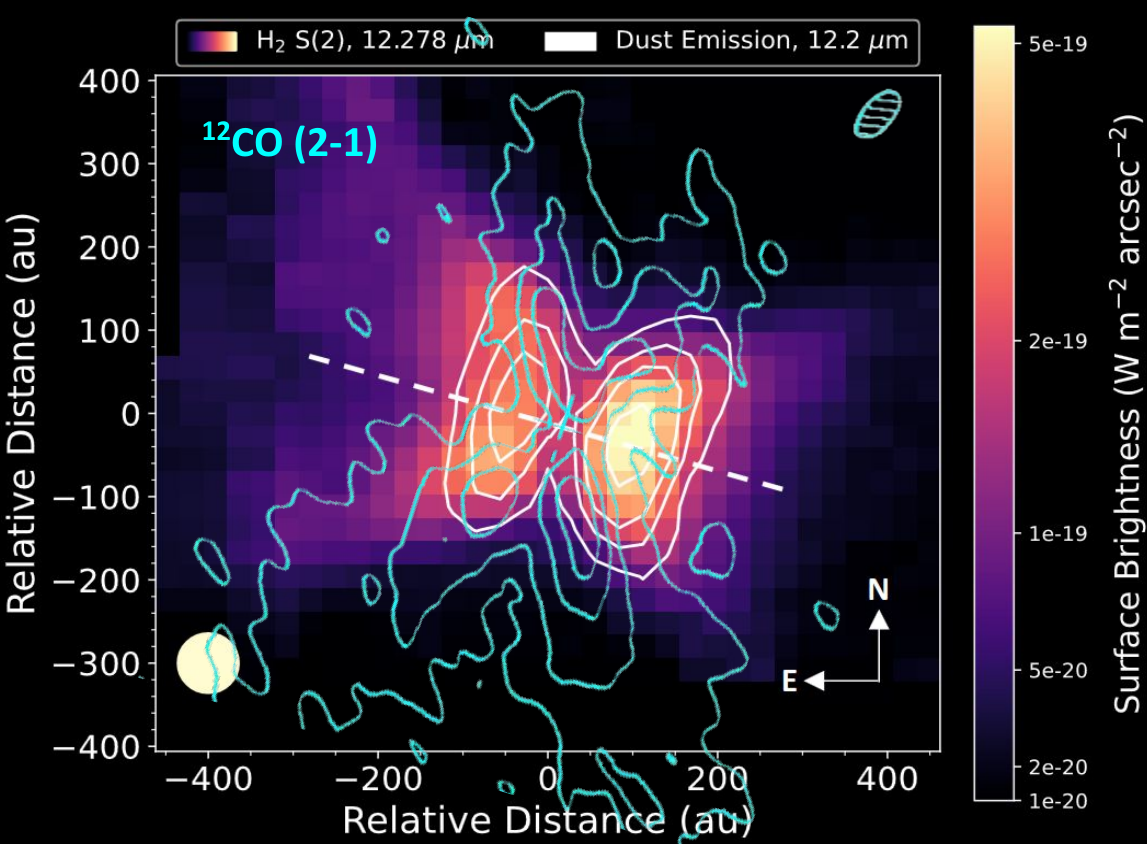}
  \centering
  \includegraphics[width=5.90cm]{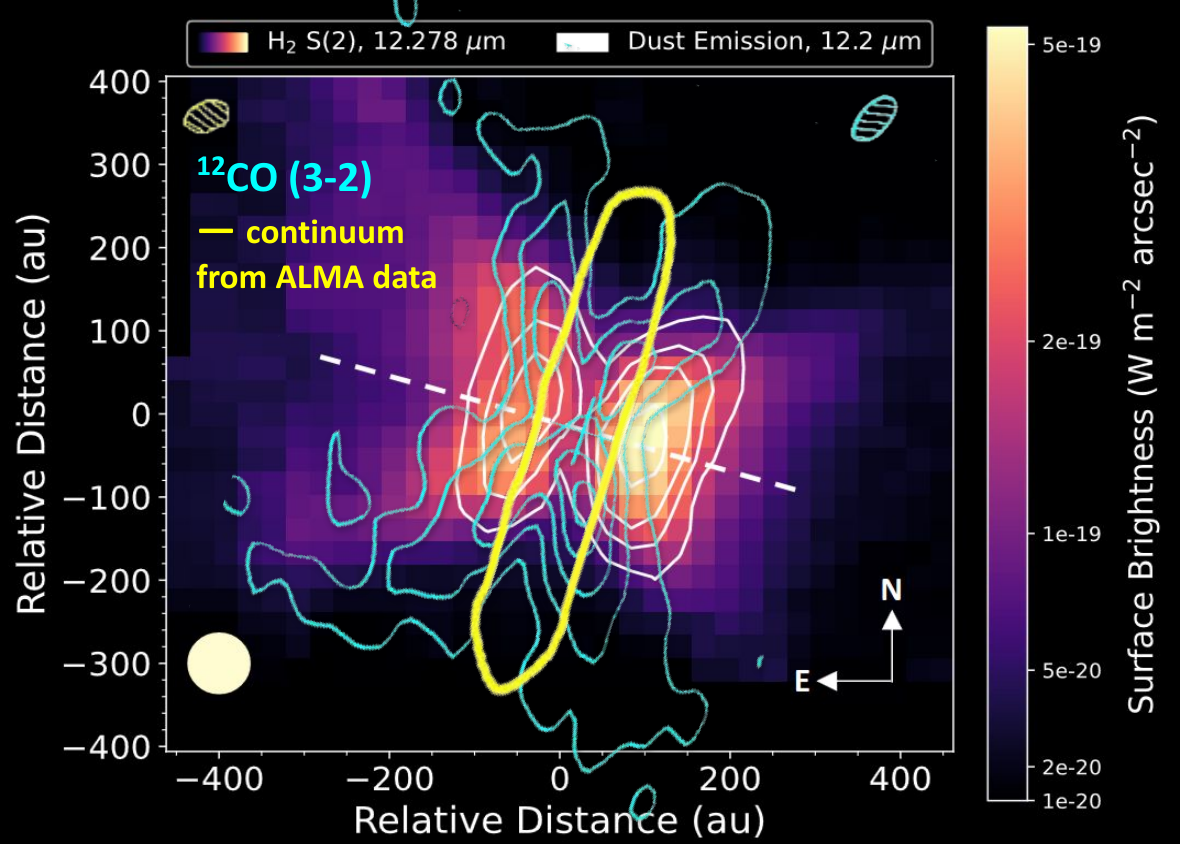}
  
  \caption{Montage of JWST and interferometric data for SSTTau042021. MIRI MRS images from JWST reveal H$_{2}$ S(2) emission at 12.278 $\mu$m (\cite{Arulanantham+2024}). White contours indicate the mid-infrared scattered-light continuum at 12.2 $\mu$m, and the white dashed line marks the rotation axis of the disk at PA = 73.5$^\circ$.
  From archival ALMA data and new NOEMA observations, cyan contours show molecular emission lines of HCO$^+$ 3–2, $^{13}$CO 2–1, C$^{18}$O 2–1, $^{12}$CO 2–1, and $^{12}$CO 3–2, while yellow contours represent the 0.85 mm continuum emission at 1 mJy/beam.  Beige circles in the bottom left corner of each panel represent the average theoretical FWHM of the PSF for JWST data, calculated from the relationship reported in \cite{Law+2023}, and the beam of the molecular line emission is shown as a cyan ellipse in the top-right corner of each panel. The contours for HCO$^+$ (3–2) range from 0.02 to 0.2 Jy/beam.km/s with a step of 0.02 Jy/beam.km/s, i.e 2.5\,$\sigma$; for $^{13}$CO (2–1), they range from 0.02 to 0.08 Jy/beam.km/s with a step of 0.01 Jy/beam.km/s, i.e 3.5\,$\sigma$; for C$^{18}$CO (2–1), they range from 0.008 to 0.016 Jy/beam.km/s with a step of 0.008 Jy/beam.km/s, i.e 3.5$\,\sigma$; for $^{12}$CO (2–1), from 0.01 to 0.08 Jy/beam.km/s with a step of 0.02 Jy/beam.km/s, i.e 6\,$\sigma$; and for $^{12}$CO (3–2) , from 0.06 to 0.8 Jy/beam.km/s with a step of 0.06 Jy/beam.km/s , i.e 4.2$\sigma$.
  }
  \label{fig:jwst}
\end{frame} 
\end{figure*}

\subsection{Data Reduction}

\paragraph{Proper Motions:} ALMA and NOEMA sampled four different epochs between October 2016 and November 2024, allowing us to measure the proper motion of the source. 
The values ($\mu_\alpha = 14.6 \pm 2.2$ and $\mu_\delta = -22.2\pm 2.0$ mas/yr) are consistent with those
of the nearby T Tau stars. 
All data were recentered to the source position RA = 04:20:21.4568 and Dec 28:13:48.99 at Epoch 2016.0 for easier reference to GAIA. 

\paragraph{Imaging:}\label{sub:confusion}
All imaging was performed using  \textsc{IMAGER}\footnote{see https:imager.oasu.u-bordeaux.fr}.

For all the data, images were obtained with natural weighting, and, when needed, rotated by -73.5° to bring the disk rotation axis along the Y axis.
Left panels of Fig.\ref{fig:dust} show the images from ALMA at 146, 224.4 and 333 GHz. 
The source continuum structure, heavily resolved in one direction, precluded the use of self-calibration. 
Continuum images are thus dynamic range limited, but spectral line data are limited by thermal noise. 

Figure \ref{fig:vel_channel} shows the $^{12}$CO 2-1, $^{12}$CO 3-2, $^{13}$CO 2-1, C$^{18}$O 2-1 with the HCO$^+$ 3-2 channel maps. Integrated area maps of CO 2-1 and HCO$^+$ 3-2 are shown in Fig.\ref{fig:moments}. 
The $^{12}$CO maps show that emission is absent in some channels (from 7.11 to 8.07 km/s for the 2-1 line and 7.04 to 7.89 km/s for the 3-2 line). This results from contamination by CO emission from clouds along the line of sight. As a result, the CO moment maps (Fig.\ref{fig:moments}) exhibit an asymmetry between the blue and red-shifted sides.
Given the maximum recoverable scale (MRS, see Table \ref{tab:obs}), only little flux is lost (less than 10 $\%$). However, the recovered flux is affected by confusion with the molecular cloud in CO.

Figure \ref{fig:jwst} is a montage showing the superimposition of the CO and HCO$^+$ intensity maps with the H$_2$ S(2) emission from the JWST.

\section{Tomographies}

A tomography is a 2D image of the gas temperature as a function of radius and height. The image is obtained by averaging the observed intensity along the lines of the position-velocity (PV) diagram. Since the disk is in rotation, each of these lines corresponds to a specific radius within the disk. This geometric characteristic allows for the direct reconstruction of the two-dimensional luminosity distribution \citep{Dutrey+etal_2017}. The resulting tomographic image has a constant linear resolution with height above the disk mid-plane, while resolution varies with radius, decreasing at larger radii, due to the decrease in Keplerian shear relative to local linewidth \citep{Dutrey+etal_2017}.

The TRD, using the mean temperature for each radius, are presented in left panels of Fig.\ref{fig:tomo}. 
TRD reveal some top-bottom asymmetries which are likely due to the fact that the disk is not perfectly edge-on \citep{Dutrey+etal_2017}. Moreover, the $^{12}$CO TRD were produced using all velocity channels, and are thus affected by the confusion with the molecular clouds, making their interpretation more difficult (although this confusion is limited to areas of integration along the radii). 

TRDs are efficient to directly visualize the vertical stratification, but the radial resolution vary because of the ratio Keplerian shear to intrinsic local line width decreases with radius. Also, a direct interpretation of the TRD is only possible for optically thick, thermalized lines.  Radiative transfer modeling is thus required to properly constrain the gas properties.

\tablecontinuum

\section{Modeling}

The limited quality of the CO data due to CO cloud confusion and  a very short integration time precludes the use of a complex modeling approach. We therefore adopt a parametric radiative transfer code, the \textsc{DiskFit} tool \citep{Pietu+2007} to estimate the basic physical parameters (Tables \ref{tab:obtained-cont} \& \ref{tab:gas-model}). The DiskFit adjustment  is made by comparing the predicted visibilities to the observations in the Fourier plane.

The vertical gas density distribution follows a Gaussian profile with a scale height $H(r)$ defined by $n(r,z)=n(r,0)\exp(-(z/H(r))^2)$.
The scale height, molecular surface densities and rotational or dust temperatures are defined as power laws with radius, the disk size being truncated by sharp inner and outer radii. 

The vertical temperature profile can be taken as isothermal or assume a gradient mimicked by parametric laws \citep[routinely used for disks e.g.][]{Law+2021}. We used here the specific prescription of \citet{Dutrey+etal_2017}. 
We use the so-called "LTE mode" in DiskFit, which provides excitation temperature for the observed lines. 
The local velocity dispersion (thermal and turbulent motions) is assumed constant over the disk. 
Details are given in Appendix \ref{app:modeling}.

\subsection{Continuum data}
We analyzed the continuum data assuming the dust absorption coefficient is a power law of frequency with a value $0.052$\,cm$^2$g$^{-1}$ at 345 GHz \citep[including a gas-to-dust ratio of 100,][]{Beckwith+1990}. 
The dust emissivity index $\beta$ was also a free parameter in the minimization process and converged towards a value of 0.6.  
We derived the geometrical parameters and then minimized the dust surface density  $\Sigma_{D}$(r) and temperature T$_D$(r). 
Because dust emission is concentrated around the disk mid-plane we assumed a simple radial power law with a vertically isothermal profile. 
The data sets were first independently minimized (the same geometrical parameters were found for the three data sets). The dust scale height $H_d(r)$ was derived from 220 and 330 GHz data only. 
We then minimized the three ALMA data sets (continuum at 146, 255 and 333 GHz), together and Table \ref{tab:obtained-cont} presents the resulting best fit model. At 146 GHz, the residuals reach about $0.1\,\sigma$, while they drop to $6.55 \times 10^{-3}\,\sigma$ at 225 GHz and $1.33 \times 10^{-2}\,\sigma$ at 333 GHz.
The larger residuals at 146\,GHz (see Fig.\ref{fig:dust}) may indicate a more settled dust layer of larger grains that are not well represented at the other frequencies.

\tablemodel

\subsection{Molecular lines}
In the analysis of the $^{12}$CO 2-1 and 3-2 lines, channels contaminated by cloud emission were ignored (see Sec. \ref{sub:confusion} and Fig.\ref{fig:vel_channel}). We modeled the line data with and without continuum subtraction, and obtained similar results for geometrical and velocity parameters, as well as for the temperature law. The final fit was performed with the continuum, using the best fit dust parameters from Table \ref{tab:obtained-cont}. Given that molecular depletion induced by freeze-out may occur in the disk mid-plane owing to the very low temperatures derived from dust, we also verified that the fitted gas temperature law was similar whether assuming or not  freeze-out (following two criteria, one on the gas temperature T$(r,z)$ and a second on the H$_2$ column density, see Appendix \ref{app:modeling} for details).

Finally, because of its low signal-to-noise, we analyzed the C$^{18}$O 2-1 emission  assuming the $^{13}$CO parameters and only fitted the surface density at 100 radius. 

Table \ref{tab:gas-model} presents the best model parameters for the five lines $^{12}$CO 2-1 and 3-2, $^{13}$CO 2-1, C$^{18}$O 2-1 and HCO$^+$ 3-2. Figure \ref{fig:tomo} presents the TRD of the observations (left panels), the TRD of the best result DiskFit models (middle panels) and their differences (right panels: observation TRD minus model TRD). Figure \ref{fig:distrib} shows the two dimensional density and temperature distributions derived for $^{12}$CO, $^{13}$CO, HCO$^+$ and the dust. 

\figtomo 

\section{Discussion}

\subsection{Dust disk}
The best fit model (Table \ref{tab:obtained-cont}) shows that the dust disk has no cavity, is large and relatively cold with a radially constant temperature of $\sim$10 K. Millimeter observations being sensitive to large grains, potentially colder than the smaller ones \citep{Gavino+etal_2021}, such a low temperature is therefore not surprising. Within the error bars, the dust temperature is also of the order of the gas temperature (8-10\,K, Table \ref{tab:gas-model}). We searched for possible variations of $\beta$ with radius \citep[as observed by ][]{Guilloteau+etal_2011,Testi+etal_2014,Perez+2012} and we found that the best fit is given for a constant value of $\beta$. 
 
The right panel of Fig. \ref{fig:dust} shows the map of the difference between the observations and the best models for the three frequencies. The model appears more centrally peaked than the observations, i.e. less optically thick, which is unusual for observations of an edge-on disk. Nevertheless, this part of the disk being strongly beam diluted, these results need to be taken with caution. Moreover, since we were not able to self-calibrate the data, some low level features seen around the disk on the left panels (particularly at 1\,mm) may be due to uncorrected small phase errors. However, Fig.\ref{fig:dust}, panels c suggest evidences for a gap at radius $\sim$70-150 au (on both sides), particularly at 150 and 330 GHz. 
However, this feature is seen neither in CO nor in HCO$^+$. More molecular data are needed to investigate the origin of such a potential gap.

The surface density (given by $\Sigma_D(r)= \Sigma_{D0}(r/100\,\mathrm{au})^{-p}$) has a power law index $p$ which is very flat ($0.5\pm0.2$).  
\citet{Bohler+2013} showed that when settling of large grains onto the mid-plane occur, the fitted exponent $p$ can become negative. 
The derived value suggests an intermediate case with moderate settling, perhaps because of a relatively
young age like for the Butterfly Class I system \citep{Grafe+etal_2013}, or possibly also because settling is less efficient due to the very low mass of the star ($0.27\Msun$).   

By integrating the dust surface density, we estimated the total mass (gas+dust) of disk to be $\sim 4.6\times10^{-2} \Msun$ (assuming a gas-to-dust ratio of 100). 
This value is consistent with the lower limit reported by \citet{Villenave+etal_2020}.

\subsection{A disk wind traced by CO?}

Figures \ref{fig:vel_channel}, \ref{fig:jwst}, and \ref{fig:tomo} show that the CO emission exhibits a larger vertical extent compared to the other molecules.

\begin{figure}[h]
\begin{frame}
\centering
 \includegraphics[width=8cm]{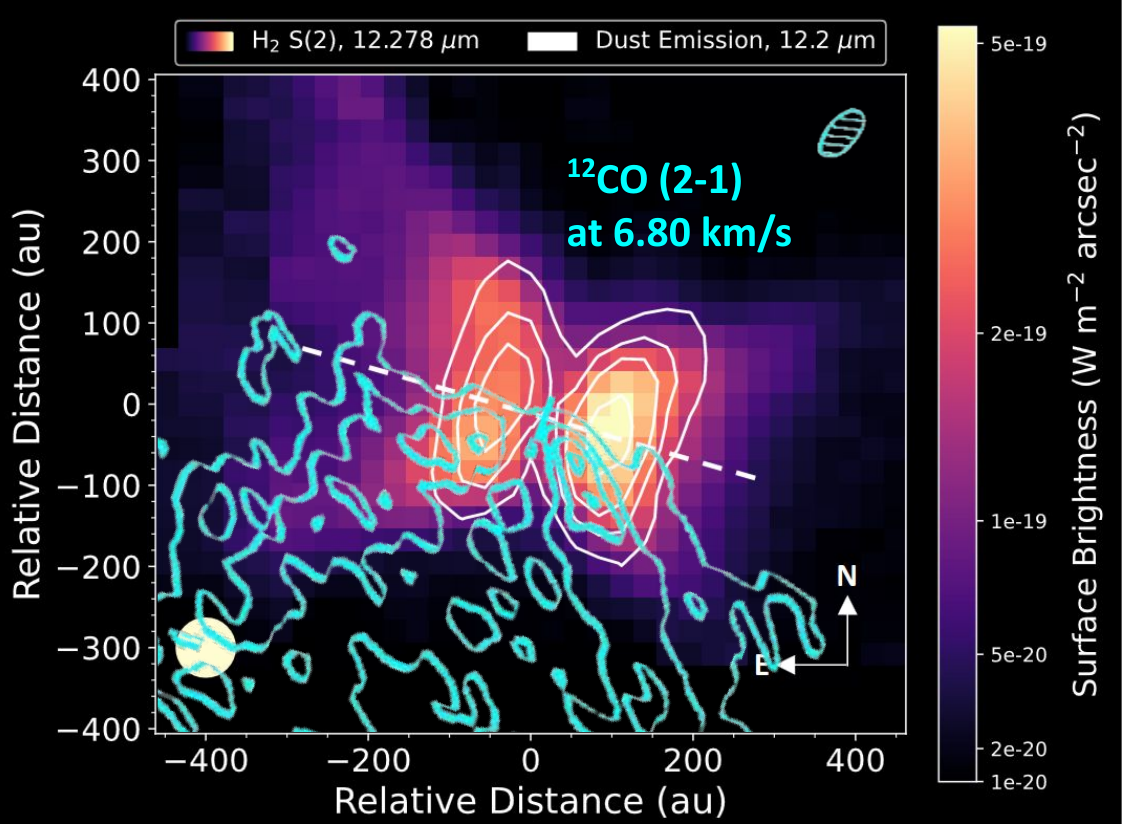}
  \centering
  \includegraphics[width=8cm]{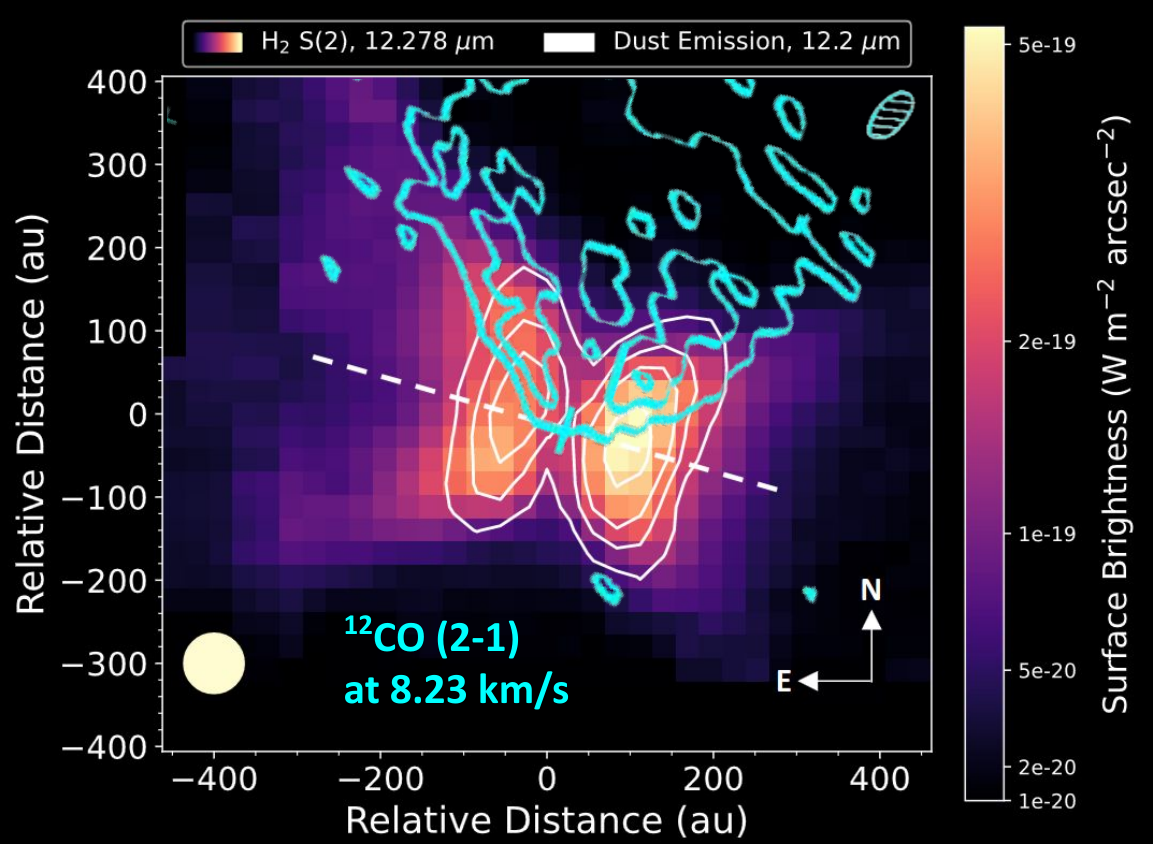}
  \centering
  \caption{ As in Figure \ref{fig:jwst}, overlay of the JWST image from \cite{Arulanantham+2024} with contours of the $^{12}$CO (2-1) channel maps at 6.80 km/s and 8.23 km/s illustrating the large vertical extent around the H$_2$ emission.
  The contours are in step of $3.4\sigma$ and range from 10 to 40 mJy beam$^{-1}$.}
  \label{fig:jwst-CO}
\end{frame} 
\end{figure}

Figure \ref{fig:jwst-CO} presents two specific channels of the  $^{12}$CO 2-1 emission that are less affected by the foreground cloud superimposed on the H$_2$ S(2) integrated area map from \citep{Arulanantham+2024}.

As can be seen in Fig.\ref{fig:jwst-CO}, the two large torii of CO gas encircle the H$_2$ S(2) disk wind launching from the top and bottom of the disk atmosphere, at above $\sim4$ scale heights. Such nested 
morphology of the disk molecular winds has been predicted theoretically \citep{Wang+2019}
and recently been confirmed by the JWST observations of several edge-on disks by \cite{Pascucci+2025}. Molecules like CO can be launched by the wind from the upper disk molecular layer high up into the heavily FUV-irradiated disk atmosphere, where they get gradually dissociated and cannot be easily produced again via ion-molecule reactions due to very low densities and long corresponding collisional timescales \citep{Panoglou+2012}. The vertical distribution of CO appears to be more confined to the disk compared to the H$_2$ IR emission due to either (1) less sensitivity of ALMA at sub-mm compared to the high sensitivity of JWST at mid-IR wavelengths and/or (2) more efficient UV dissociation of the CO gas, despite its self-shielding and mutual shielding by the H$_2$, as CO molecules are much less abundant than H$_2$ molecules \citep{VanDishoeck+1988}.

On the other hand, there is no significant signature of such an extended morphology in the $^{13}$CO, C$^{18}$O and HCO$^+$ emission, as can be seen from Figs.\ref{fig:vel_channel} and \ref{fig:jwst}. This would be expected in a disk wind interpretation because of the insufficient column densities of these molecules in the disk atmosphere, where the disk wind launches from. The relative faintness in CO 3-2 may
also indicate a low density or low temperature there.

We note that in our DiskFit analysis, we excluded velocity channels contaminated by the cloud emission (as indicated by the crosses in Fig.\ref{fig:vel_channel}) which cover a substantial part of the putative disk wind region. Thus, this disk wind should not affect our disk parameters determination.

The main issue affecting the interpretation is contamination from the surrounding molecular cloud, rather than the disk wind itself.

\subsection{Gas temperature}

Table \ref{tab:gas-model} and Figs.\ref{fig:tomo}, \ref{fig:distrib} present the results of the best models for the observed molecules. Note that the results from $^{12}$CO may be partly corrupted by the confusion with the CO clouds and that due to beam dilution, our results are not significant below a radius of $\sim$50 au. 

At radius around 50-250 au, i.e. in the densest disk area, we found that the atmospheric gas temperature is essentially larger for $^{12}$CO than for $^{13}$CO. This is not surprising since the bulk of the $^{13}$CO gas is located at a lower altitude than the much more optically thick $^{12}$CO emission which reflects more the whole extent of the molecular gas. With or without freeze-out in the disk mid-plane, we always derived a temperature for the molecular layer of 15-20K, very close to the freeze out temperature of CO.

Determining the mid-plane temperature is more tricky because it a priori depends on the assumed
shape of the vertical temperature profile, and may also be affected by the continuum emission.

However, all models converge towards low mid-plane temperature values around 9 K at 100 au. 
 
This temperature rises  then to 25 K in the upper layers. 
Similar temperatures are also observed for the edge-on disk around the Flying Saucer by \citet{Dutrey+etal_2017}. Our values are lower than those obtained by \citet{Flores+etal_2021} for the brighter source Oph1631 that is also observed edge-on. They found about 20 K at a radius $<$ 200 au near the mid-plane, and a temperature of $\sim$30 K for the upper layer, with an almost vertically isothermal profile beyond a radius of 200 au.

The (excitation) temperature distributions derived from HCO$^+$ 3–2 and $^{13}$CO 2–1 closely agree. Since the later transition is thermalized,
this implies the HCO$^+$ 3-2 line is thermalized too setting up a lower limit to the density in the emitting region.

We tried to assess the impact of our simple mathematical prescription onto the results (Appendix \ref{app:modeling}). The best fit model (which has a value of the exponent $\delta\simeq 2.5-3$ that controls the steepness) suggests that the transition of the temperature from the mid-plane to the bulk of the molecular emission region, which is at lower altitude for the  $^{13}$CO and HCO$^+$ than for $^{12}$CO, is very steep.

\subsection{Molecular distribution}

In the mid-plane, the TRD (Fig.\ref{tab:obtained-cont}, panels (a)) suggest an absence of emission. This may be a result of the lower gas temperature in the mid-plane, as found for HD\,163296 by \citet{Dullemond+2020}. 
However, our extrapolated mid-plane temperature is much lower than the 18\,K found in the above case, so that we also expect molecules to be under-abundant at these very low temperatures due to freeze-out on cold grains
\citep{Gavino+etal_2021, Ruaud+2019}.
We handled this effect using models with a complete freeze-out of molecules in cold and UV shielded regions (see Appendix \ref{app:modeling} and Fig.\ref{fig:distrib} for details). This improves the fits, particularly for the CO (2–1) model, by reducing the residuals near the mid-plane (see Fig.\ref{fig:depletion}), which suggests that such a depletion is indeed present. As shown in the figure \ref{fig:depletion}, when there is no depletion in the model, the residuals near the disk mid-plane increase by approximately 1K. Although this effect is less pronounced in the residual maps of the other lines, the data quality for those transitions is not sufficient to clearly confirm this behavior. Nevertheless, since the effect is clearly observed for $^{12}$CO, it is more consistent to assume that a similar depletion is present for all lines. 

Several studies have provided direct evidence of molecular gas extending beyond the dust disk's outer radius in protoplanetary disks. For instance, \citet{Panic+2009} and \citet{Cleeves+2016} observed $^{12}$CO emission in the IM Lup disk extending out to 970\,au, while the millimeter continuum emission is truncated at 313\,au. 
Despite a more limited signal to noise, our observations also indicate the presence of cold molecular gas beyond the dust disk's outer edge. 
The \textsc{DiskFit} analysis clearly confirms

\noindent
this,
with much larger outer radii for CO and HCO$^+$ than for dust.

In addition, the TRD of the residuals for $^{13}$CO and C$^{18}$O seem to exhibit some extra emission beyond the dust outer radius (at $>350-400$ au), suggesting an additional gas component that is not represented by our
simple power disk model.

Vertically, the molecular extents of the HCO$^+$ and $^{13}$CO emission appear relatively similar. The TRD of the differences (Fig. \ref{fig:tomo}, panel (c)) show for $^{13}$CO that above and below the mid-plane, the model also exhibits excess emission which remains however difficult to quantify because of the limitations imposed by the data quality. We note however that the apparent height of HCO$^+$ is about 30\% larger than that of $^{13}$CO. 
The molecular scale heights, 24 to 30\,au at 100 au, should not be
confused with the hydrostatic equilibrium scale height, about 15\,au given the mid-plane temperature of 10 K, neither to the dust scale height whose value (11\,au) suggests moderate settling.
This suggests that HCO$^+$, the most abundant molecular ion in disks, is essentially formed at an intermediate altitude in the disk (at about 2 hydrostatic scale heights).  This is in agreement with the work from \citet{vanthoff+etal_2017} who utilized detailed physical and chemical modeling to demonstrate that HCO$^+$ is predominantly formed in the warm molecular layer of the disk, specifically at intermediate heights above the mid-plane. 

Finally, since HCO$^+$ 3-2 seems to be thermalized, we can determine a lower value for the local H$_2$ density at radius 100 au, which should be significantly higher than the critical collision density of the 3-2 rotational HCO$^+$ transition $\sim2.2\times10^{6}$\,cm$^{-3}$ \citep{Denis-Alpizar+2020}.
Given the scale height at which the emission is visible, this translates to $\sim10^8$\,cm$^{-3}$ in the mid-plane, a value consistent with the density extrapolated from dust emission.

\section{Summary}
\begin{itemize}
    \item[$\bullet$] CO clouds complicate the analysis, but the $^{12}$CO emission is vertically extended, partly tracing the H$_2$ wind observed by JWST.
    \item[$\bullet$] HCO$^+$, C$^{18}$O and $^{13}$CO trace the bulk of the molecular layer.
    \item[$\bullet$] The gas and dust mid-plane temperatures are $\sim$8-12\,K, the molecular layer temperature (from $^{13}$CO and HCO$^+$) is $\sim$16\,K while the CO gas atmospheric temperature is $\sim$31\,K at 100 au.
    \item[$\bullet$] HCO$^+$ 3-2 is thermalized, setting a lower limit for H$_2$ density of $\sim2.2\times10^{6}$\,cm$^{-3}$ at 100-200 au.
    \item[$\bullet$] CO and HCO$^+$ TRD residuals suggest that some gas is on the mid-plane beyond the dust outer radius ($\geq$ 300 au).
\end{itemize}

\begin{acknowledgements}
This work is based on observations carried out under projects number W22BC \& S24AT with the IRAM NOEMA Interferometer [30m telescope]. IRAM is supported by INSU/CNRS (France), MPG (Germany) and IGN (Spain).
 This work was supported by the Thematic ActionS ``Programme National de Physique Stellaire'' (PNPS) and “Physique et Chimie du Milieu Interstellaire” (PCMI) of INSU Programme National “Astro”, with contributions from CNRS Physique \& CNRS Chimie, CEA, and CNES.
This research made use of the SIMBAD database, operated at the CDS, Strasbourg, France.
Y.W.T. acknowledges support through NSTC grant 111-2112-M-001-064- and 112-2112-M-001-066-.
This paper makes use of the following ALMA data: 2016.1.00771.S and 2016.1.00460.S.
ALMA is a partnership of ESO (representing its member states), NSF (USA) and NINS (Japan), together with NRC (Canada), NSTC and ASIAA (Taiwan), and KASI (Republic of Korea), in cooperation with the Republic of Chile. The Joint ALMA Observatory is operated by ESO, $AUI/NRAO$ and NAOJ.
This work was also supported by the NKFIH NKKP grant ADVANCED 149943 and the NKFIH excellence grant TKP2021-NKTA-64. 
N. T. Phuong acknowledges support from Vietnam Academy of Science and Technology under grant number VAST 08.02/25-26"
\end{acknowledgements}

\bibliographystyle{aa}
\bibliography{ref-abaur.bib}

\begin{appendix}
\onecolumn
\section{Observations}
\begin{figure*}[!b]
    \hspace*{-10mm}
    \includegraphics[width=18.7cm]{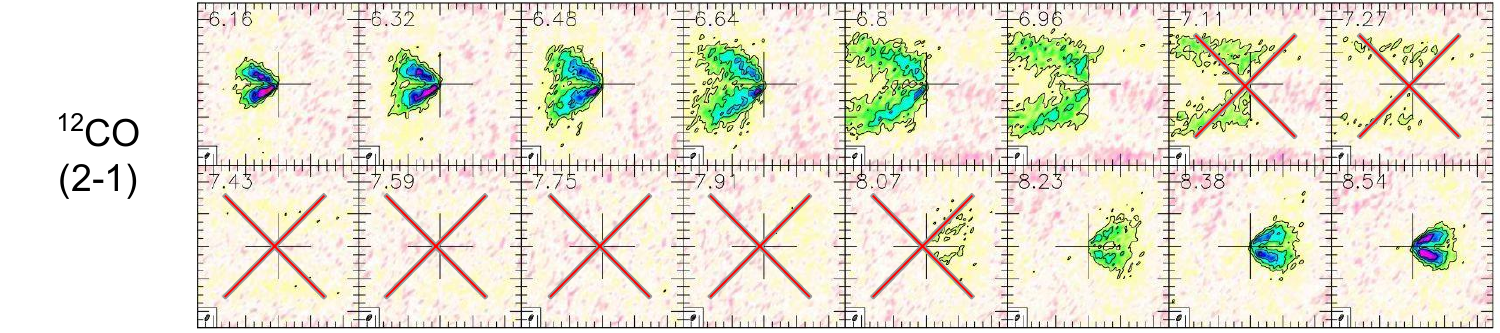}
    \hspace*{-10mm}
    \includegraphics[width=18.7cm]{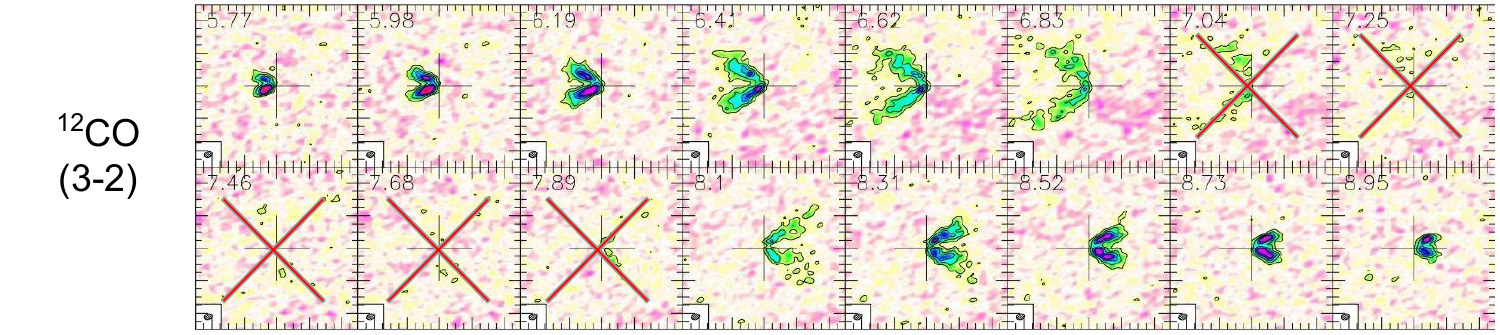}
    \hspace*{-10mm}
    \includegraphics[width=18.8cm]{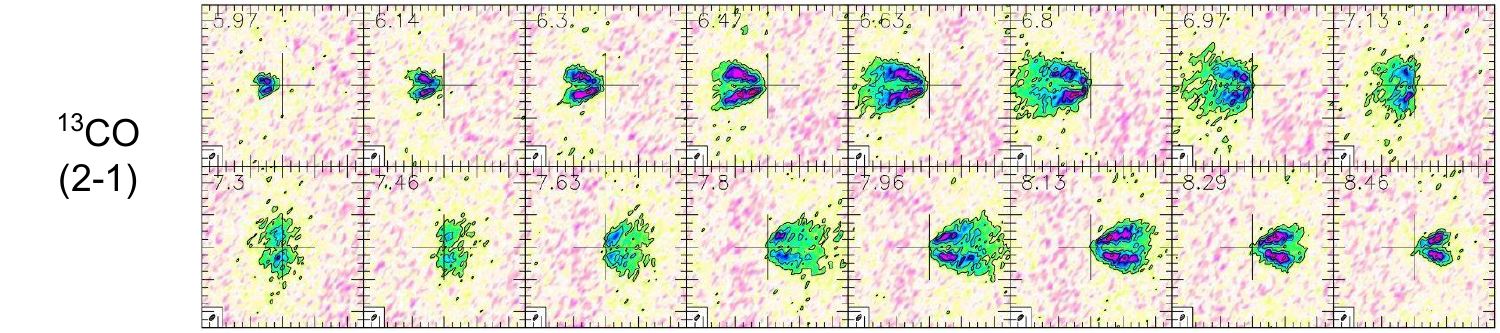}
    \hspace*{-10mm}
    \includegraphics[width=18.75cm]{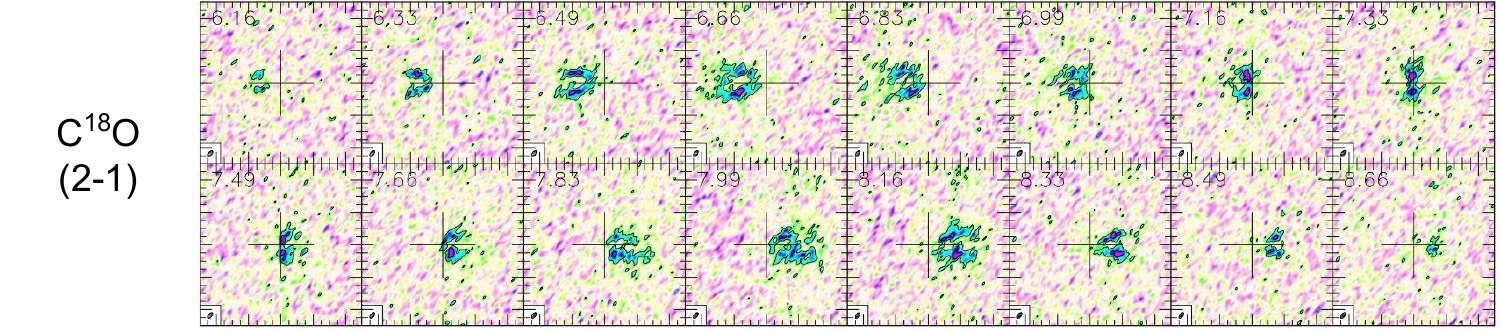}
    \hspace*{-4.5mm}
    \includegraphics[width=18.8cm]{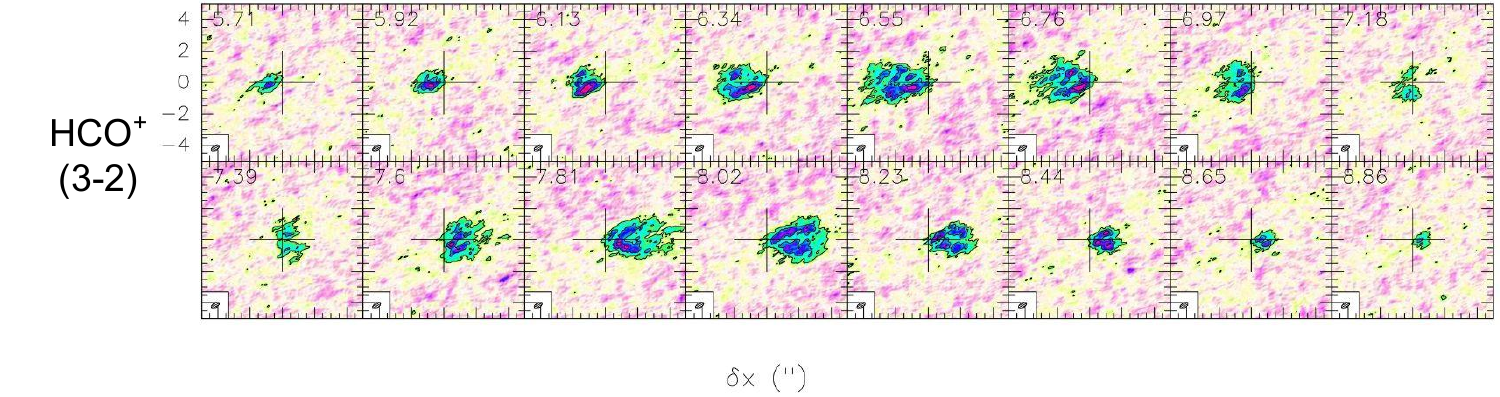}

    \caption{ Velocity channel maps from top to bottom: for $^{12}$CO 2-1, 3-2, $^{13}$CO 2-1, C$^{18}$O 2-1 and HCO$^{+}$ 3-2. The channels corrupted by the CO clouds along the line of sight are marked with a red cross. These channels were not used in the \textsc{DiskFit} analysis. All the maps are rotated by $-73.5^\circ$. The contours are defined as follows: from 3 to 12 K in steps of 3 K (i.e., $3.6\sigma$) for the $^{12}$CO 2–1, $3.3\sigma$ for $^{12}$CO 3–2, and $3.6\sigma$ for $^{13}$CO 2–1; from 2 to 6 K in steps of 2 K (i.e., $3.2\sigma$) for C$^{18}$O 2–1; and from 3 to 9 K in steps of 3 K (i.e., $3.3\sigma$) for HCO$^+$ 3–2.}
    \label{fig:vel_channel}
\end{figure*}

\twocolumn
\section{Modeling}
\label{app:modeling}
\subsection{Basic parameters}

In \textsc{Diskfit} \citep{Pietu+2007}, a disk is characterized by several parameters, for which error bars can be derived. These parameters include the atmospheric and mid-plane temperatures ($T_{atm}$,$T_{mid}$, the scale height ($H(r)$), the surface density ($\Sigma(r)$), the inner and outer radius (R$_{in}$,R$_{out}$), disk velocity,
disk orientation and inclination, and stellar mass. For the final minimization, we fix the pivot R$_0$ at 100 au to maximize the sensitivity. 
The version of the model we used operates under the assumption of LTE conditions and azimuthal symmetry. For lines which are not thermalized, the derived temperature thus corresponds to the line excitation temperature.
The vertical distribution of the matter is assumed to follow a Gaussian profile, as $n(r,z) = n(r,z=0) \exp(-(z/H(r))^2)$, where the scale height $H(r)$ varies as a power law with radius.
Note that with such a definition of scale heights, the hydrostatic scale height in the disk mid-plane would
be $H = \sqrt{2}C_s/\Omega_K$ where $C_s$ is the isothermal sound speed and $\Omega_K$ the Keplerian angular velocity,
a factor $\sqrt{2}$ larger than the alternate definition used in some other works.

The local velocity dispersion $\delta V$, which encompasses thermal and turbulent motions on scales smaller than the beam size, is assumed to be radially constant, as the data have insufficient sensitivity to reveal possible spatial variations.
To find the best model by minimization, \textsc{Diskfit} proceeds by comparison in the $uv$ plane in order to reduce the non linear deconvolution effects.

\subsection{Depletion Zone}
To constrain the presence of gas onto the mid-plane, the final runs were performed including a depletion factor near the mid-plane, with two criteria, one onto the gas temperature $T(r,z)$ and a second onto the H$_2$ column density from the current height to the outside, $\Sigma_Z(r,z) = \int_z^\infty n(r,u) du$. 
 
Molecules are only present if $T(r,z) > T_{dep}$ (with accounts for molecules sticking on grains at low temperatures) or $\Sigma_Z(r,z) < \Sigma_{dep}$ (to account for photo-desorption from dust when sufficient interstellar UV radiation penetrates).

Following the results reported by \cite{Qi+2013}, we assumed a
depletion temperature of 17.5\,K for all our molecules.
We found similar values for $\Sigma_{dep} \approx 3.6\,10^{22}$ cm$^{-2}$
for CO and $^{13}$CO, while for HCO$^{+}$, $\Sigma_{dep}$=$8.7\pm0.2\,10^{22}$ cm$^{-2}$ suggests that HCO$^+$ extends closer to the mid-plane than CO.

\begin{figure*}[htbp!]
    \centering
    \includegraphics[width=16cm]{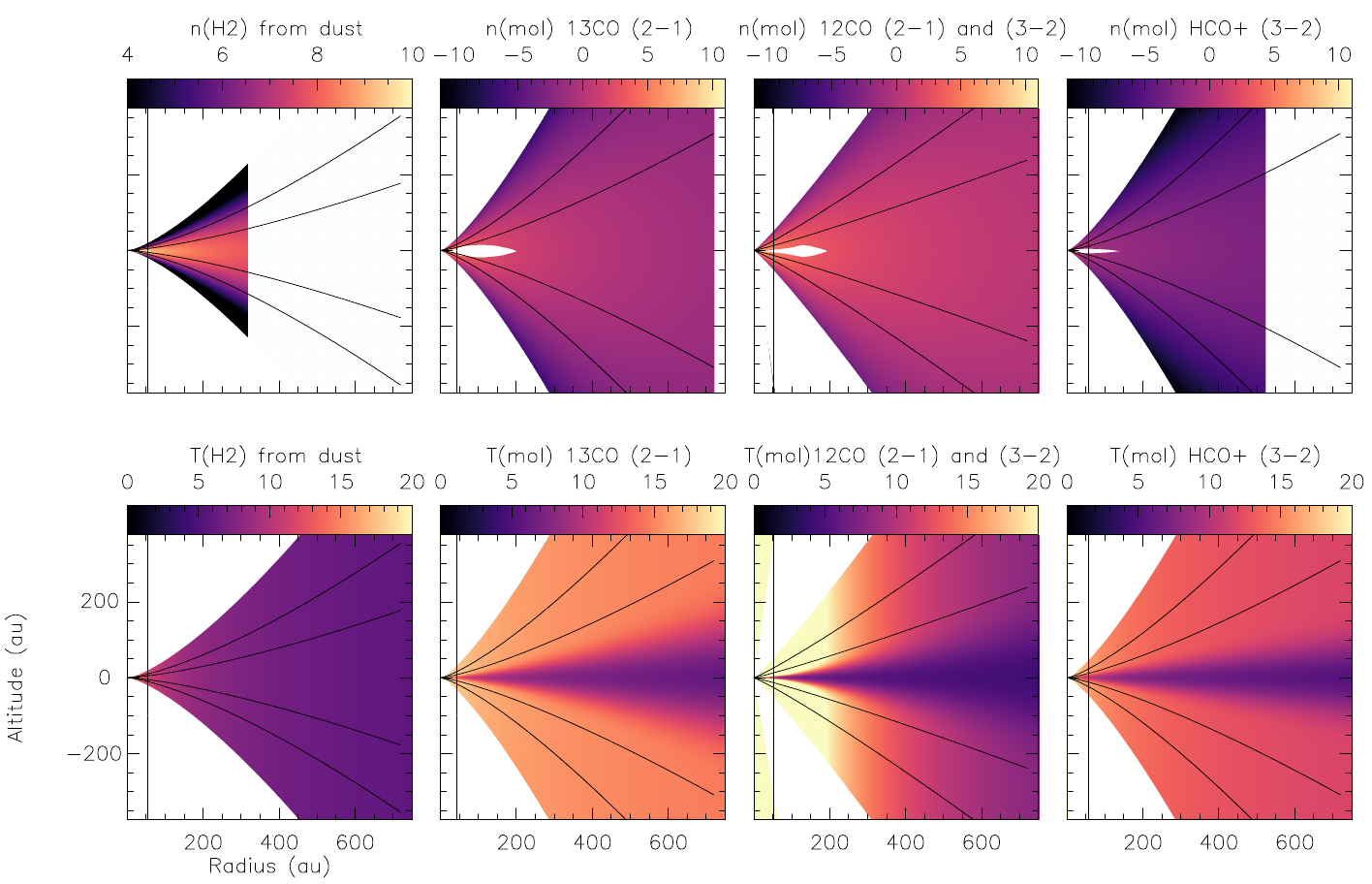}
    \caption{Density profile on the first row and temperature profile on the second, for the DiskFit best models. The 1st column is for dust, the 2nd for $^{13}$CO, the 3rd for $^{12}$CO (2-1) and (3-2) and the 4th for HCO$^+$. The solid vertical line represents the beam, and the curved horizontal lines correspond to one and two scale heights. Note: the density profile scale (in cm$^{-3}$) is expressed in $log_{10}$. The densities shown for the molecular tracers correspond to the number density of the specific molecule, not the total H$_2$ density, and are therefore several orders of magnitude lower than dust-derived gas densities.}
    \label{fig:distrib}
\end{figure*}

\begin{figure*}[htbp!]
    \centering
      \centering
  
    \subfigure{\includegraphics[width=0.45\textwidth]{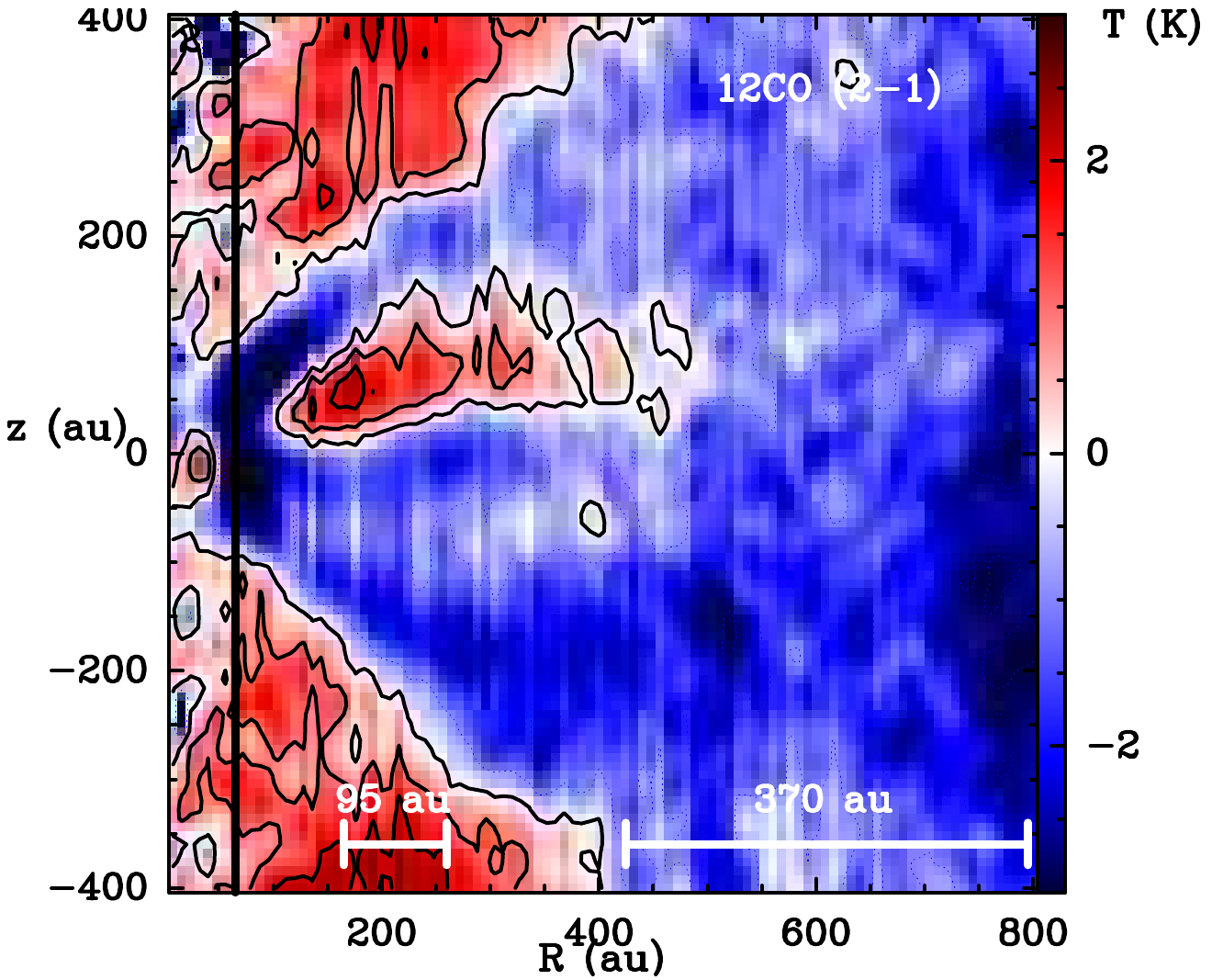}} 
  \subfigure{\includegraphics[width=0.45\textwidth]{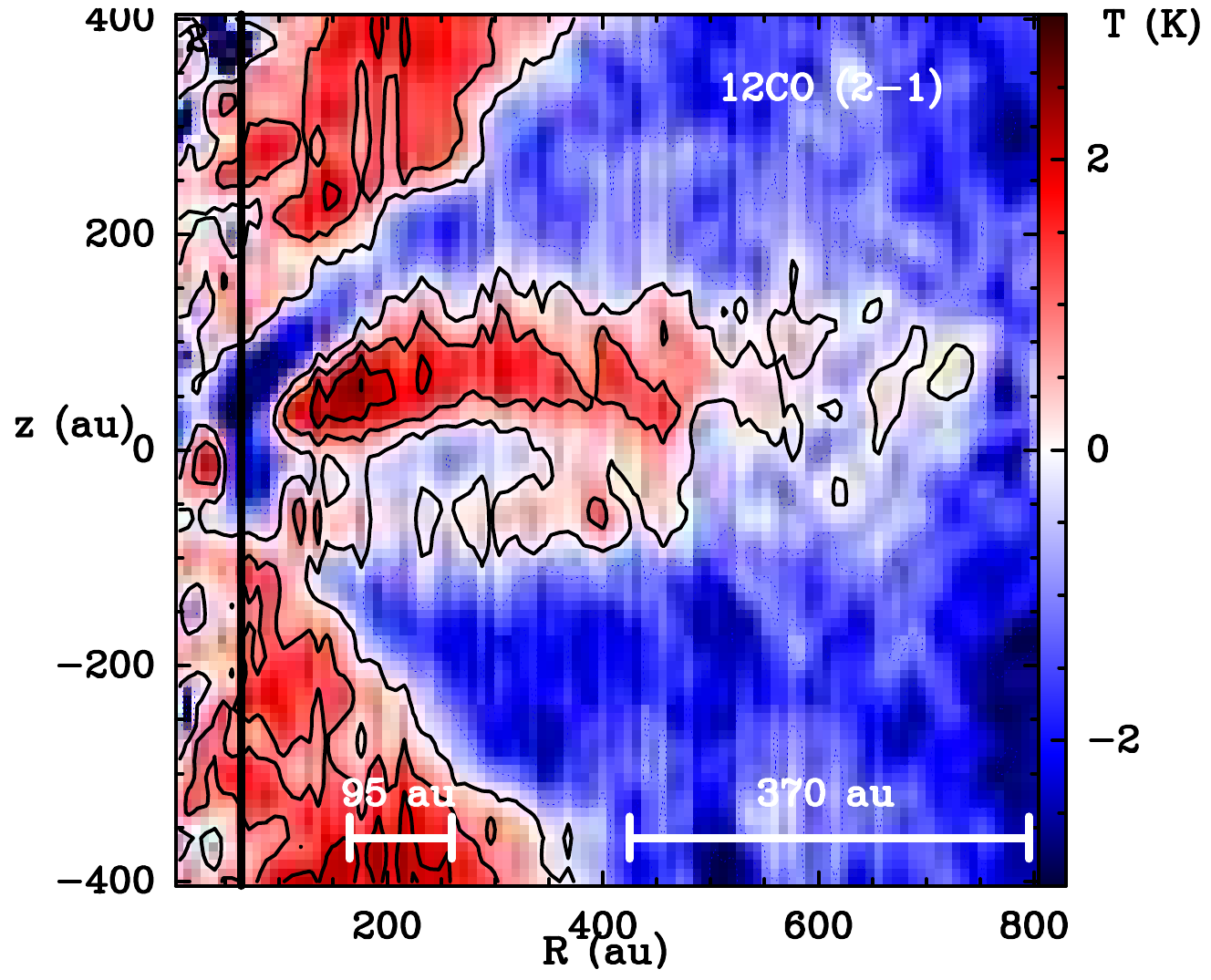}}

  \caption{Differences TRD (observation TRD - model TRD) of the $^{12}$CO 2-1 emission after fitting a model with depletion (on the left) and without depletion (on the right). 
   Color scale spans and contours are defined from -3\,K to 3\,K in steps of 1\,K. }

    \label{fig:depletion}
\end{figure*}

The better fits obtained with depletion is illustrated in Fig.\ref{fig:depletion} which shows the residual TRD of $^{12}$CO with and without depletion zone.  
For HCO$^+$, which actually has a smaller depletion area (see Fig.\ref{fig:distrib}),  the difference is less significant, as for 
$^{13}$CO and C$^{18}$O because of their limited opacities.
 
\subsection{Temperature profile}
Temperature radial and vertical profiles are modeled following \citet{Dartois+etal_2003} and \citet{Dutrey+etal_2017}. 
The disk atmosphere temperature is given by: 
\begin{equation}
    \label{Atmosphere temperature}
    T_{atm}(r) = T^{0}_{atm}\left(\frac{r}{r_0}\right)^{-q_{atm}}
\end{equation}
while the mid-plane temperature corresponds to: 
\begin{equation}
    \label{Mid-plane temperature}
    T_{mid}(r) = \mathrm{min}\left(T_{atm}(r),T_{0}\left(\frac{r}{r_0}\right)^{-q_{0}}\right)
\end{equation}

In between, for an altitude of $z < z_qH(r)$, the temperature is defined by:
\begin{equation}
    \label{Temperature}
    T(r,z) = \left(T_{mid}(r)-T_{atm}(r)\right)\left(\cos\left(\frac{\pi z}{2z_qH(r)}\right)\right)^{2\delta}+T_{atm}(r)
\end{equation}

The atmospheric temperature was minimized together with the values of $z_q$ and $\delta$ which controls the 
position and sharpness ($\delta$) of the vertical temperature gradient.
We did several checks to determine their impact on the fit. The best fit models systematically correspond to a very steep value of $\delta\simeq 2.5-3$, but simultaneous
adjustment of $\delta$ and $z_q$ were not possible. We finally opted for $\delta = 2.5$
and found $z_q \sim 1.2$, indicating that the steep gradient is at about $0.3\,H(r)$. 
Here, $H(r)$ is the apparent scale height of the molecule distributions: the hydrostatic scale height derived from the mid-plane temperature is about a factor 2 lower.\\

We tested the robustness of our results by modifying the functional form of the temperature profile and replacing the cosine dependence with a sine function. This alternative prescription leads to nearly identical results in the minimization procedure, both in terms of temperature structure and line emission fits. 

\begin{equation}
    \label{eq:temperature}
    T(r,z) = \left(T_{mid}(r)-T_{atm}(r)\right)\left(\sin\left(\frac{\pi z}{2z_qH(r)}\right)\right)^{2\delta}+T_{atm}(r)
\end{equation}

\begin{table}[!th] 
\small
\caption{Best fit models for $^{13}$CO 2-1 assuming Eq.\ref{eq:temperature}}       
\centering    
\begin{tabular}{|c c|l|l|l|}        
\hline              
\multicolumn{2}{|c|}{Parameters} &  \multicolumn{1}{c|}{$^{13}$CO 2-1} \\ 
\hline  
$\delta$V & (km\,s$^{-1}$) &  $0.20 \pm 0.01$ \\
$T_{atm}(r)$ & (K) & $ 15.1 \pm 0.4\,(r/R_0)^{-0.32 \pm 0.11}$ \\
$T_{mid}(r)$ & (K)&  $\min(T_{atm},11.1 \pm 0.3\, (r/R_{0})^{-0.28})$ \\
$\Sigma(r)$ & (cm$^{-2}$) & $3.3 \pm 0.2 \, 10^{17}\,(r/R_{0})^{-3.2 \pm 0.2}$ \\ 
$\Sigma_{dep}$   & (cm$^{-2}$) & $3.7\pm0.4\,10^{22}$   \\
$\delta$ &  & 2.5 $\pm$ 0.4 \\
$z_q$  &  & 1.4 $\pm$ 0.3 \\
\hline 
Inclination & ($^\circ$) & $89.8 \pm 0.6$ \\
R$_{in}$ & (au) &  50 $\pm$ 1 \\
R$_{out}$ & (au) & 720 $\pm$ 5 \\
 \hline 
\end{tabular}
\label{tab:gas-model-bis}
\end{table} 

This consistency demonstrates that our conclusions are not strongly dependent on the specific parametric form adopted for the vertical temperature gradient, and that our analysis captures the essential physical structure of the disk. Furthermore, \cite{Galloway+2025} used several parametric functions to analyse data from exoALMA and found that the function in cosine provided a better fit.

\end{appendix}

\end{document}